\documentstyle[preprint,prd,aps]{revtex}

\begin{document}


\draft

\title{Post-Newtonian Models of Binary Neutron Stars}

\author{James C.~Lombardi, Jr.}
\address{Center for Radiophysics and Space Research,
         Cornell University, Ithaca, NY 14853}
\address{Department of Astronomy, Cornell University, Ithaca, NY 14853}
\author{Frederic A.~Rasio}
\address{Department of Physics, MIT 6-201, Cambridge, MA 02139}
\author{Stuart L.~Shapiro}
\address{Departments of Physics and Astronomy, University of Illinois at
Urbana-Champaign, 1110 West Green Street, Urbana, IL 61801}
\address{National Center For Supercomputing Applications, University of
Illinois at Urbana-Champaign, Urbana, IL 61801}


\maketitle


\begin{abstract}
Using an energy variational method, we calculate quasi-equilibrium
configurations of binary neutron stars modeled as compressible triaxial
ellipsoids obeying a polytropic equation of state.  Our energy
functional includes terms both for the internal hydrodynamics of the
stars and for the external orbital motion.  We add the leading
post-Newtonian (PN) corrections to the internal and gravitational
energies of the stars, and adopt hybrid orbital terms which are
fully relativistic in the test-mass limit and always accurate to PN
order.  The total energy functional is varied to find quasi-equilibrium
sequences for both corotating and irrotational binaries in circular
orbits.  We examine how the orbital frequency at the innermost stable
circular orbit depends on the polytropic index $n$ and the compactness
parameter $GM/Rc^2$.  We find that, for a given $GM/Rc^2$, the
innermost stable circular orbit along an irrotational sequence is about
17\% larger than the innermost secularly stable circular orbit along
the corotating sequence when $n=0.5$, and 20\% larger when $n=1$.  We
also examine the dependence of the maximum neutron star mass on the
orbital frequency and find that, if PN tidal effects can be neglected,
the maximum equilibrium mass increases as the orbital separation
decreases.
\end{abstract}

\pacs{04.25.Nx,04.30.Db,04.40.Dg,97.60.Jd}

\narrowtext

\section{Introduction}

The mergers of neutron star, black hole, and neutron star-black hole
binaries provide the most promising sources of gravitational waves for
detection by laser interferometers now being constructed, such as LIGO
and VIRGO.  A variety of additional factors also motivate the study of
binary neutron star systems:  for example, such systems are known to
exist with several having been detected in our Galaxy in the form of
binary radio pulsars (eg. \cite{nst96} and references therein), and
coalescing neutron stars may be a source of gamma-ray bursts at
cosmological distances \cite{npp92}.  These facts have motivated
numerous theoretical studies of the inspiral and coalescence of compact
binaries.  Until recently, such calculations have involved either
post-Newtonian (PN) point mass approximations or purely Newtonian
hydrodynamic treatments.

Analytic analyses of the gravitational radiation waveforms have been
performed in the PN approximation with the binary components, whether
black holes or neutron stars, being treated as point masses
\cite{stu123,kww,bdiww95}.  These investigations are therefore
appropriate in the early inspiral phase when the stars are still well
separated and tidal effects can be safely neglected.  Semi-analytic
\cite{k92,bc92,LRS1,LRS2,LRS3,LRS4,LRS5,ls95} and numerical
\cite{stu8910} studies of the coalescence waveform from binary neutron
stars also have been performed using Newtonian hydrodynamics and the
quadrupole radiation approximation.  These studies treat Newtonian
effects arising from the finite size of the interacting fluid stars and
focus on their tidal distortion, disruption, and eventual merger.
Numerical simulations of coalescing black hole and neutron star
binaries are also underway in full general relativity (eg.
\cite{stu11,wmm}).  Complicated by the challenges of treating both
matter and strong gravitational fields in (3+1) dimensional spacetime,
these simulations are essential for obtaining definitive, quantitative
results for the highly relativistic interactions which generate the
late inspiral and final coalescence waveforms.

There is an important distinction between binaries in Newtonian theory
and those in general relativity:  In Newtonian theory a binary can
exist in true dynamic equilibrium, but in general relativity true
equilibrium cannot exist because of the generation of gravitational
waves.  For binaries with sufficiently large orbital separation ($r >>
M$), the orbital decay time due to emission of gravitational waves is
much longer than the orbital timescale, so that the inspiral is
quasi-static through a sequence of quasi-equilibrium configurations.
The inspiral becomes dynamic and quasi-equilibrium is destroyed with
the onset of relativistic orbital instabilities ($r\sim$ few $M$).
Dynamic behavior may also commence when the purely Newtonian tidal
instability is reached near $r\sim$ few $R$ (eg.,
\cite{LRS1,LRS2,LRS3,LRS4,stu8910}).  For these reasons, both Newtonian
and relativistic effects are important.

In the relativistic simulations of Wilson, Mathews, and Marronetti
\cite{wmm}, the time-dependent equations of relativistic hydrodynamics
are solved in a simplified, strong-field background, and the system is
relaxed to find quasi-static equilibrium configurations of corotating
neutron star binaries for a specific equation of state.  These authors
claim that neutron stars just below the maximum (isolated) mass become
unstable and collapse to black holes when placed in close binary
orbits, in contrast to the Newtonian result that the tidal field tends
to stabilize a star (see Fig. 15 of Ref.  \cite{LRS1}; also Ref.
\cite{l96}).  Such collapse could significantly affect the inspiral and
the associated gravitational waveform, and therefore deserves careful
attention.

Recently, Baumgarte et al. \cite{b97} have constructed quasi-static
binary equilibrium models in full general relativity.  However, their
numerical models, based on a polytropic equation of state, are
restricted to be corotating.  This work complements the numerical
analysis of Cook \cite{c94}, who analyzed binary black holes with zero
spins.

A fully general relativistic calculation of a binary neutron star
configuration is a computationally intensive
problem.  By contrast, an ellipsoidal figure of equilibrium
(EFE) treatment, while only approximate, can find an equilibrium
configuration in less than a second on a typical workstation.  This
speed affords a quick means of gaining a qualitative understanding of
the stellar models and equilibrium sequence.  These sequences can
help provide a better understanding of the general relativistic
solutions which are now being generated (eg. Ref. \cite{b97}).  In
addition, while general relativistic codes are currently limited to
corotation, or near corotation, an EFE treatment allows straightforward
construction and comparison of both corotating and irrotational
sequences.  The second order variation of the functional can be used to
identify instabilities, e.g. the innermost stable circular orbit (ISCO)
or the collapse of a massive neutron star in a binary to a black hole.
Furthermore, both secular and
dynamical instabilities can be identified and distinguished.

Taniguchi and Nakamura \cite{tn96} use the EFE approach to model black
hole-neutron star binaries in the Roche approximation, using a
pseudo-potential to model the strong-field orbital motion.  Lai and
Wiseman \cite{lw96} insert EFE terms in the
hybrid-P$^2$N dynamical equations of motion, which are then integrated
to identify the ISCO.  The EFE approach has also recently been extended
to first PN order in treatments of corotating binaries \cite{ts97}.
        
In this paper, we adopt the EFE energy variational method, extended to
first PN order. Our energy functional includes terms for both the
internal hydrodynamic structure as well as the external orbital
motion.  We add the leading PN correction to the internal and
self-gravitational energy of the stars.  We adopt the hybrid
expressions of \cite{kww}, to PN order, to give the orbital terms;
these expressions are exact in the test mass limit and correct to PN
order for arbitrary masses.  The energy functional is varied to find
quasi-equilibrium configurations of both corotating and irrotational
binary sequences, parameterized by orbital separation (or orbital
angular velocity).

In \S\ref{approximations} we outline the basic
approximations of our EFE scheme.  In \S III we give the various
energy, angular momentum, and circulation terms which comprise our
functionals.  The method of solution is discussed in \S IV, and
results are presented in \S V.  Finally, we discuss the implications of
our results and directions for future work in \S VI.  In the appendices, we
derive the PN energy contributions for an isolated spherical star obeying a
polytropic equation of state.

\section{Fundamental Approximations}\label{approximations}

In this section, we describe the fundamental approximations and
assumptions behind our energy variational method, which we apply to
construct general Darwin-Riemann equilibrium models.  Further details
and applications of our method can be found in
\cite{LRS1,LRS2,LRS3,LRS4}.

In steady state, an isolated, self-gravitating binary system is
characterized by conserved global quantities such as the rest
(baryonic) mass $M_o$ and $M_o'$ of each star and the total angular
momentum $J$.  The total energy of the system can be written as a
functional of the rest mass density and velocity distributions
$\rho_o({\bf x})$ and ${\bf v}({\bf x})$, respectively, even if the
system is not in equilibrium. In principle, an equilibrium
configuration can be determined by extremizing this energy functional
with respect to all variations of $\rho_o({\bf x})$ and ${\bf v}({\bf
x})$ that leave the conserved quantities unchanged. The fundamental
approximation of our method is to replace the infinite number of
degrees of freedom contained in $\rho_o({\bf x})$ and ${\bf v}({\bf
x})$ by a finite number of parameters $\gamma_1,~\gamma_2,\ldots$, so
that the total energy is a function only of these parameters:
\begin{equation}
E=E(\gamma_1,\gamma_2,\ldots;\, M_o, M_o', J, \ldots). \label{lrs4.2.1}
\end{equation}
An equilibrium configuration is then
determined by extremizing the energy according to
\begin{equation}
{\partial E \over \partial \gamma_i}=0,~~~i=1,2,\ldots \label{lrs4.2.2}
\end{equation}
where the partial derivatives are taken holding the conserved
quantities $M_o,\,M_o',\,J,\ldots\,$ constant.

An expression like equation~(\ref{lrs4.2.1}) can be written down only
with the help of simplifying assumptions.  We adopt the compressible
ellipsoidal approximation to model the neutron stars, i.e. we assume
they are triaxial ellipsoids supported in hydrostatic equilibrium by a
polytropic equation of state \cite{k92,LRS1,LRS2,LRS3,LRS4,LRS5,c69}.
Surfaces of constant density are assumed to be concentric self-similar
ellipsoids.  These approximations become exact in the limit that the
fluid is incompressible and the gravitational potential is strictly
Newtonian and truncated beyond the tidal term \cite{c69}.  For
compressible configurations, this simplification is equivalent to using
an ellipsoidal trial function in an energy functional and then
extremizing the functional to find the equilibrium solution.
Comparisons with numerical simulations show that the ellipsoidal model
is reliable, particularly for stars obeying stiff equations of state
\cite{LRS1,LRS2,LRS3,LRS4}.

The velocity field of the fluid is taken to be that of a Riemann-S
ellipsoid, in which all internal velocities are linear functions of the
coordinates (see eg. Ref. \cite{LRS1}, \S5, for details).  The vorticity
vector is assumed to be everywhere parallel to the orbital rotation
axis.  Since the viscosity in neutron stars is thought to be
negligible, the two stars will conserve circulation as they inspiral
\cite{k92,bc92}, so that slowly spinning neutron stars at $r=\infty$
will maintain a small, constant circulation.  However, current general
relativistic quasi-equilibrium codes are forced to consider corotating,
or nearly corotating, sequences rather than sequences of constant
circulation.  We therefore concentrate on two interesting limiting
cases: irrotational sequences (for non-synchronized systems with zero
circulation) and corotating sequences (corresponding to synchronized
binary systems in uniform rotation).

Consider a binary system composed of two stars of rest mass $M_o$
and $M_o'$ in circular orbit.
Throughout this paper unprimed quantities refer to the star of rest mass
$M_o$ while primed quantities refer to the star of mass $M_o'$. 
Following Ref. \cite{c69}, we denote the mass ratio as $p\equiv M_o/M_o'$. 
The density and pressure are related by
\begin{equation}
P=K\rho_o^{(1+1/n)},~~~~P'=K'\rho_o'^{\,(1+1/n')}. \label{lrs4.2.3}
\end{equation}
Note that our treatment allows the two stars to have both distinct
polytropic indices ($n\ne n'$) and polytropic constants ($K\ne K'$),
although in this paper we concentrate on the case of two identical
stars.  The binary separation is denoted by $r$, and the principal axes
of the two ellipsoids by $a_1$, $a_2$, $a_3$, and $a_1'$, $a_2'$,
$a_3'$. The orientation is such that $a_1$ and $a_1'$ are measured
along the binary axis, $a_2$ and $a_2'$ in the direction of the orbital
motion, and $a_3$ and $a_3'$ along the rotation axis.  In place of the
three principal axes $a_i$, it is often convenient to introduce as
independent variables the central rest mass density $\rho_{oc}$, and
two oblateness parameters defined as
\begin{equation}
\lambda_1 \equiv \biggl({a_3\over a_1}\biggr)^{2/3},~~~~
\lambda_2 \equiv \biggl({a_3\over a_2}\biggr)^{2/3}. \label{lrs4.2.4}
\end{equation}
Similarly we can introduce $\rho_{oc}'$, $\lambda_1'$, and $\lambda_2'$
in place of the three $a'_i$.  Thus the seven independent variables
which parameterize our models are
$\{r,\,\,\rho_{oc},\,\lambda_1,\,\lambda_2,\,\,\rho_{oc}',\,\lambda_1',\,\lambda_2'\}$.

As is standard in EFE treatments, we assume that the density profile
$\rho(m)$, where $m$ is the mass interior to an isodensity surface, is
identical to that of a spherical polytrope with the same $K$ and $n$,
but with radius $R=(a_1a_2a_3)^{1/3}$  [see eq. (\ref{ov-based})].  In
particular, to PN order, a star with semi-major axes $a_i$ has the same
central density as a spherical star (not necessarily in equilibrium)
with radius $R$ in Schwarzschild coordinates,
\begin{equation}
\rho_{oc} = {M_o\xi_1\over 4\pi R^3|\theta_1'|}\left\{
1
+{GM_o\over Rc^2}{1\over n+1}
\left[
      {3\alpha_1\over\xi_1^2|\theta_1'|^2}
      +{\beta_1\over\xi_1|\theta_1'|^2}
      +{(n-3)(n+1)\over 5-n}
\right]
\right\}, \label{rhoofr}
\end{equation}
where we have used equations (\ref{surface}) and (\ref{final.aofm}).
Here $\theta$ and $\xi$ are the usual Lane-Emden variables for a
polytrope (see, e.g., Ref. \cite{c39}) with $\theta(\xi_1)=0$, and
where the quantities $\alpha_1$ and $\beta_1$ are defined in
Appendix~\ref{PN.expand.OV} and depend only on the polytropic index $n$
(for $n=0.5$ we have $\alpha_1=-3.39307$ and $\beta_1=3.63708$, while
for $n=1.0$ we have $\alpha_1=-2.77960$ and $\beta_1=2.67168$).  In the
Newtonian limit ($c\rightarrow\infty$), equation~(\ref{rhoofr}) reduces
to the expression used in the Newtonian treatment of
\cite{LRS1,LRS2,LRS3,LRS4}. Although equation~(\ref{rhoofr}) originates
from a PN expansion, our code treats this relation as exact.

\section{Energy, Angular Momentum, and Circulation Terms} \label{ejcterms}

\subsection{Newtonian Treatment} \label{Newtonianefe}

We now review the energy, angular momentum, and circulation
terms in the Newtonian EFE approximation, which has been covered in detail in
\cite{LRS1,LRS2,LRS3,LRS4}.  When similar expressions can be written
for the two stars, we give only the one corresponding to $M_o$, the
other being obtained simply by replacing unprimed with primed
quantities.

The usual Newtonian orbital contributions to the energy and angular
momentum are
\begin{equation}
E_{N\,orb}={1\over 2}\mu \Omega^2 r^2-{GM_oM_o'\over
r},~~~~J_{N\,orb}=\mu r^2\Omega, \label{Newtonianorbitalej}
\end{equation}
where $\mu=M_oM_o'/(M_o+M_o')$ is the reduced mass and $\Omega=v/r$ is the
orbital angular velocity.  The gravitational tidal (quadrupole)
interaction energy $W_t$ between the two stars is (see Appendix~B of
Ref. \cite{LRS1})
\begin{equation}
W_t=-{GM_o\over 2r^3}(2I_{11}'-I_{22}'-I_{33}')
-{GM_o'\over 2r^3}(2I_{11}-I_{22}-I_{33}). \label{lrs4.2.27}
\end{equation}
Here we have defined 
\begin{equation}
I_{jj}={h_j(\lambda_1,\lambda_2)\over
4k_3}M_o^{5/3}\rho_{oc}^{-2/3}~~~~\hbox{(no summation)},
\label{lrs1.B8}
\end{equation}
where
\begin{equation}
h_1\equiv{\lambda_2\over\lambda_1^2},~~~~h_2\equiv{\lambda_1\over\lambda_2^2},~~~~h_3\equiv\lambda_1\lambda_2,\label{lrs1.B9}
\end{equation}
with analogous relations for $I_{jj}'$ and $h_j'$.  We will refer to
key relations in Ref. \cite{LRS1} by their equation number preceded by
an ``I''; for example, equations~(\ref{lrs1.B8}) and (\ref{lrs1.B9})
are also equations~(I.B8) and (I.B9), respectively.  The dimensionless
structure coefficient $k_3$ appearing in equation~(\ref{lrs1.B8}) is
defined by [cf.\ eq.~(I.3.17)]
\begin{equation}
k_3={5\over 4\kappa_n}\left({4\pi|\theta_1'|\over\xi_1}\right)^{2/3},
\end{equation}
where
\begin{equation}
\kappa_n\equiv{5\over3}\,{\int_0^{\xi_1}\theta^n\xi^4\,d\xi\over
\xi_1^4|{\theta'}_1|}, \label{kappan}
\end{equation}
so that $\kappa_n=1$ for $n=0$.  Values of $k_3$ and $\kappa_n$ for
various $n$ are tabulated in Table I of \cite{LRS1}.

The internal energy of the star with mass $M_o$ is given simply by 
\begin{equation}
U=k_1K\rho_{oc}^{1/n}M_o. \label{lrs4.2.5}
\end{equation}
The self-gravitational energy can be written [cf.\ eq.~(I.4.6)]
\begin{equation}
W=- k_2GM_o^{5/3}\rho_{oc}^{1/3} f,\label{lrs4.2.6}
\end{equation}
where the dimensionless ratio $f$ is given by
\begin{equation}
f=f(\lambda_1,\lambda_2)\equiv {A_1a_1^2+A_2a_2^2+A_3a_3^2 \over 2
(a_1a_2a_3)^{2/3}},\label{lrs4.2.7}
\end{equation}
so that $f=1$ for a spherical star.
The index symbols $A_i$ are defined as in Ref. \cite{c69} (\S17),
\begin{equation}
A_i\equiv a_1a_2a_3\int_0^{\infty}\!\!{du \over \Delta (a_i^2+u)},~~~~
{\rm with}~~~~\Delta^2=(a_1^2+u)(a_2^2+u)(a_3^2+u), \label{lrs4.2.8}
\end{equation}
and are functions only of $\lambda_1$ and $\lambda_2$.  In
equations~(\ref{lrs4.2.5}) and~(\ref{lrs4.2.6}), $k_1$ and $k_2$ are
dimensionless structure constants depending only on the polytropic
index $n$, defined as
\begin{equation}
k_1\equiv{n(n+1)\over5-n}\,\xi_1|{\theta'}_1|,~~~~
k_2\equiv{3\over5-n}\,\biggl({4 \pi
|{\theta'}_1|\over\xi_1}\biggr)^{1/3}. \label{k1k2}
\end{equation}
Values of
$k_1$ and $k_2$ are tabulated in Ref. \cite{LRS1}.

We permit the star to be differentially rotating with uniform vorticity
$\zeta {\bf e_3}$ as measured in the corotating frame of the binary:
\begin{equation}
\zeta \equiv ( {\bf \nabla} \times {\bf u}) \cdot {\bf e_3} =
-{a_1^2+a_2^2 \over a_1 a_2}\Lambda.  \label{lrs4.2.11}
\end{equation}
Here the velocity field ${\bf u}$ in the corotating frame is given by
\begin{equation}
{\bf u}=Q_1x_2{\bf e_1}+Q_2x_1{\bf e_2} \label{lrs4.2.12}
\end{equation}
with
\begin{eqnarray}
Q_1 & = & -{a_1^2 \over a_1^2+a_2^2} \zeta = + {a_1 \over a_2} \Lambda,
\nonumber \\
Q_2 & = & +{a_2^2 \over a_1^2+a_2^2} \zeta = - {a_2 \over a_1} \Lambda.
\label{lrs4.2.13}
\end{eqnarray}
The quantity $\Lambda$ is the angular velocity of the internal fluid
motions in the corotating frame.  Here $\bf e_1$ is along the binary
axis, directed from $M_o$ to $M_o'$, $\bf e_2$ is in the direction of
the orbital velocity, and $\bf e_3$ is perpendicular to the orbital
plane.  The velocity field in the inertial frame is given by
\begin{equation}
{\bf u}^{(0)}={\bf u} + {\bf \Omega} \times {\bf x}, \label{lrs4.2.14}
\end{equation}
and the vorticity in this frame is
\begin{equation}
\zeta^{(0)}=({\bf \nabla} \times {\bf u}^{(0)})\cdot {\bf
e}_3=(2+f_R)\Omega, \label{lrs4.2.15}
\end{equation}
where
\begin{equation}
f_R={\zeta\over\Omega}. \label{lrs4.2.10}
\end{equation}
Corotation (synchronization) corresponds to
$f_R=\zeta=\Lambda=0$, while an irrotational velocity field is
obtained when $f_R=-2$.  Note that the geometric outer shapes of the
two stars always rotate at the orbital angular velocity $\Omega$ in
the inertial frame, regardless of the internal velocity field.

The Newtonian spin kinetic energy $T_s$ (i.e., the kinetic energy in
internal fluid motions) in the inertial frame can be obtained from
equations~(\ref{lrs4.2.12})--(\ref{lrs4.2.14}).  Assuming a Lane-Emden
trial density function, one finds [cf.\ eq.~(I.5.6)]
\begin{equation}
T_s={1\over2}I(\Lambda^2+\Omega^2)
-2\left(I_{11}I_{22}\right)^{1/2}\Lambda\Omega, \label{lrs4.2.16}
\end{equation}
where $I=I_{11}+I_{22}$
is the moment of inertia. 
Similarly, the Newtonian spin angular momentum $J_s$ can be written
[cf.\ eq.~(I.5.5)]
\begin{equation}
J_s =I\Omega -2\left(I_{11}I_{22}\right)^{1/2}\Lambda. \label{lrs4.2.19}
\end{equation}

As given by \cite{LRS1,LRS2,LRS3,LRS4}, the purely Newtonian total
energy of the system, not necessary in equilibrium, is therefore
\begin{equation}
E_N=E_{N\,orb}+U+U'+W+W'+T_s+T_s'+W_t, \label{Newtoniane}
\end{equation}
Similarly, the total angular momentum to Newtonian order is
\begin{equation}
J_N=J_{N\,orb}+J_s+J_s'. \label{Newtonianj}
\end{equation}

Another important quantity, conserved in the absence of viscosity, is
the fluid circulation $C$ along the
equator of the star (see \S 5.1 of Ref. \cite{LRS1}).  We define
\begin{equation}
{\cal C}\equiv \biggl (-{1\over5\pi}\kappa_nM_o\biggr)\,C
= I\Lambda-2\left(I_{11}I_{22}\right)^{1/2}\Omega. \label{lrs4.2.20}
\end{equation}
The quantity ${\cal C}$ has the dimensions of angular momentum but is
proportional to the conserved circulation $C$.  For convenience, we
typically refer to ${\cal C}$ itself as the circulation.

\subsection{Post-Newtonian Treatment} \label{PNefe}

In this paper, we augment the Newtonian terms of \S\ref{Newtonianefe}
with PN contributions (a) to the orbital energy and angular momentum,
(b) to the internal and self-gravitational energy of a star, and (c) to
the coupling energy and angular momentum between the internal structure
and the orbital motion.

We replace of the Newtonian contributions to orbital energy and angular
momentum [eq. (\ref{Newtonianorbitalej})] with the hybrid-PN
expressions of \cite{kww} (with $\dot{r}=0$,
valid for circular orbits):
\begin{equation}
E_H=E_T+E_\eta,~~~~J_H=J_T+J_\eta
\end{equation}
where the Schwarzschild, or test-mass, contributions are (with $G=c=1$
for these two equations only)
\begin{eqnarray}
E_T & = & \mu\left\{ \left({r-m \over r+m}\right)^{1/2}\left[1-v^2{(r+m)^3
\over r^2(r-m)}\right]^{-1/2}-1\right\}, \\
J_T & = & \mu vr\left\{\left[{(r+m)^5 \over
r^4(r-m)}\right]^{1/2}\left[1-v^2{(r+m)^3 \over
r^2(r-m)}\right]^{-1/2}\right\},
\end{eqnarray}
and the PN corrections unaccounted for by the Schwarzschild terms are
\begin{eqnarray}
E_\eta & = & -\eta\mu v^2\left({9\over 8}{v^2\over c^2}-{1\over
2}{Gm\over rc^2}\right), \\
J_\eta & = & -\eta\mu vr\left({3\over 2}{v^2\over c^2}-{Gm\over
rc^2}\right).
\end{eqnarray}
Here $v=\Omega r$ is the orbital velocity, the total rest mass
$m=M_o+M_o'$, and the mass parameter $\eta\equiv\mu/m$.  In the
notation of Ref. \cite{kww}, $m$ also includes contributions from the
self energy of the stars and is not simply the total rest mass;
however, we account for these contributions below [see eqs. (\ref{eoi})
and (\ref{joi})] as a coupling between the internal structure and the
orbital motion.  With our definitions, the quantities $m$ and $\eta$
remain strictly constant along the sequence.
%
%
The radius $r$ is the so-called harmonic or de Donder coordinate.  When
expanded in $Gm/rc^2$, the above expressions for $E_H$ and $J_H$ give
the Newtonian terms [eq.  (\ref{Newtonianorbitalej})] to lowest order
and the correct orbital PN terms to first order.

The PN contributions to the internal and self-gravitational energy of a
spherical polytrope are derived in Appendix \ref{derive.l1.and.l2}:
\begin{eqnarray}
U_{PN} & = & -l_1{G \over c^2}K \rho_{oc}^{{1\over n}+{1\over
3}}M_o^{5/3}, \label{upn} \\
W_{PN} & = & -l_2 {G^2 \over c^2} M_o^{7/3} \rho_{oc}^{2/3}, \label{wpn}
\end{eqnarray}
where the coefficients $l_1$ and $l_2$ are defined by equations
(\ref{final.l1}) and (\ref{final.l2}), respectively, and depend only on
the polytropic index $n$.  Values for $l_1$ and $l_2$ are listed in
Table~\ref{structureconstants}.  We also include the analogous
expressions for the PN energy contributions $U_{PN}'$ and $W_{PN}'$ in
the second star.  For isolated stars, variation of the energy function
$E(\rho_{oc})=U+W+U_{PN}+W_{PN}$ does a good job of approximating the
exact equilibrium sequence obtained by integrating the
Oppenheimer-Volkoff (OV) equation, especially for soft equations of
state (see Figure~\ref{isolated.spherical}).

We neglect PN corrections due to asphericity; for small perturbations
from spherical, equations (\ref{upn}) and (\ref{wpn}) are valid to
lowest order.  To account completely for all PN corrections to the
internal and self-gravitational energy, equations (\ref{upn}) and
(\ref{wpn}) would need to be multiplied by a form factor analogous to
$f(\lambda_1,\lambda_2)$ in equation~(\ref{lrs4.2.7}).
We also neglect P$^2$N and higher order corrections to the internal and
self-gravitational energy of the neutron stars.


The final PN correction terms we include involve the coupling between
the orbital motion and the internal structure.  Effacement of internal
structure (see Ref. \cite{d87} and references therein) requires that we
substitute the effective masses $M_{eff}=M_o+E_{self}/c^2$ and
$M_{eff}'=M_o'+E_{self}'/c^2$ {\it in place of} the rest masses $M_o$
and $M_o'$ in the Newtonian equation~(\ref{Newtonianorbitalej}) if we
wish to maintain PN order accuracy.  Here the self energy of the mass
$M_o$ neutron star is $E_{self}=W+U+U_{PN}+W_{PN}$.  We have chosen to
exclude the spin kinetic energy $T_s$ from the self energy $E_{self}$
in order to ensure that our binary configurations satisfy the identity
$dE=\Omega dJ$ exactly (see \S\ref{find.eq.soln}).  We speculate that
additional higher order terms which we are neglecting would restore
this identity when we include $T_s$.   Furthermore, $T_s$ is typically
small, especially for irrotational sequences, so that we are neglecting
a term which is smaller than our order of accuracy.  Expanding to PN
order, the resulting corrections to the Newtonian point-mass energy and
angular momentum are
\begin{eqnarray}
E_{OI} & = & {E_{self}\over c^2}\left({1 \over 2} {M_o'^2 \over m^2}
{v^2 \over c^2} - {GM_o' \over r}\right), \label{eoi} \\
J_{OI} & = & {E_{self} \over c^2} {M_o'^2 \over m^2}vr, \label{joi}
\end{eqnarray}
with analogous expressions $E_{OI}'$ and $J_{OI}'$ due to the
corrections arising from the self energy of the neutron star of mass
$M_o'$.  The subscript OI serves as a reminder that these terms couple
the orbital motion to the internal structure of the neutron stars.
Although including $U_{PN}$ and $W_{PN}$ as part of $E_{self}$ in
equations (\ref{eoi}) and (\ref{joi}) amounts to keeping terms which
are formally of P$^2$N order, doing so keeps our energy functional
accurate even for stars at
$r=\infty$ but with $v>0$; excluding $U_{PN}$ and $W_{PN}$ from
$E_{self}$ would lead to an energy functional which yields manifestly
incorrect results for isolated stars (in particular, the stability of
a star would depend on the velocity of the observer).

We neglect PN effects in the internal motion which would appear as
corrections to the spin kinetic energy $T_s$, the spin angular momentum
$J_s$, and the circulation $C$.  For corotating sequences, these
corrections are on the order of $R/r$ times the PN orbital terms, so
that we are justified in neglecting PN internal motion effects all the
way down to roughly the ISCO ($r\sim$ few $R$).  For irrotational
sequences, PN internal motion is especially small, since the fluid
motion in the inertial frame is always predominately set by the orbital
motion.  We are also neglecting all PN tidal terms (but see
\S\ref{future.work}).

The total energy and angular momentum functionals used in our treatment
are therefore
\begin{equation}
E=E_H+U+U'+W+W'+T_s+T_s'+W_t+U_{PN}+U_{PN}'+W_{PN}+W_{PN}'+E_{OI}+E_{OI}',
\label{efunctional}
\end{equation}
and
\begin{equation}
J=J_H+J_s+J_s'+J_{OI}+J_{OI}'. \label{jfunctional}
\end{equation}
Since we are approximating the internal fluid motion as Newtonian, we
continue to use equation~(\ref{lrs4.2.20}) for the circulation.

\section{Finding Equilibrium Sequences}\label{find.eq.soln}

In order to find an equilibrium solution, we must extremize the energy
function $E$ while holding all the quantities $M_o$, $M_o'$, ${\cal
C}$, ${\cal C}'$, and $J$ constant.  We are therefore identifying those
configurations for which the energy is unchanged to first order by any
perturbation which conserves rest mass, circulation, and total angular
momentum.  An equilibrium sequence is constructed by repeating this
extremization at various values of the orbital separation $r$.  In the
case of corotation, the stars are members of the Riemann-S family with
vorticity $\zeta=0$, and although the perturbations considered conserve
circulation, the circulation does indeed vary along the equilibrium
sequence itself \cite{LRS1,LRS4}.

Unfortunately, the various energy terms in \S\ref{ejcterms} do not
depend explicitly on $J,~{\cal C}$, and ${\cal C}'$.  Instead, the
energy has been written in terms of the variables $\sigma_i$ from the
set
$\{r,\rho_{oc},\lambda_1,\lambda_2,\rho_{oc}',\lambda_1',\lambda_2',\Omega,\Lambda,\Lambda'\}$,
and these variables are not independent.  The Lagrange multiplier
method allows us straightforwardly to minimize the total energy
$E(\sigma_i)$ subject to the constraints that the total angular
momentum $J(\sigma_i)$ and the circulations ${\cal C}(\sigma_i)$ and
${\cal C'}(\sigma_i)$ are constant.  The equations which determine an
equilibrium configuration are
\begin{equation}
{\partial E \over \partial \sigma_i} + \xi_J {\partial J \over
\partial \sigma_i} + \xi_{\cal C} {\partial {\cal C} \over \partial
\sigma_i} + \xi_{\cal C'} {\partial {\cal C'} \over \partial \sigma_i} =
0, \label{lagrange}
\end{equation}
where $\xi_J$, $\xi_{\cal C}$, and $\xi_{\cal C'}$ are the Lagrange
multipliers.  Here the partial differentiation with respect to
$\sigma_i$ is done with all the remaining variables $\sigma_j$ ($j\neq
i$) being held constant.  Although equations (\ref{lagrange}) are
accurate only to the order of our expansion of $E$, $J$, ${\cal C}$,
and ${\cal C'}$, our numerical code treats them as exact.  We expanded
and restructured a version of the code developed in Ref.  \cite{LRS4}
to solve these coupled equations numerically.  For a given value of
$r$, we use a Newton-Raphson scheme to solve the 10 equations
(\ref{lagrange}) and find the variables
$\rho_{oc},\lambda_1,\lambda_2,\rho_{oc}',\lambda_1',\lambda_2',\Omega,\xi_J,
\xi_{\cal C}$, and $\xi_{\cal C'}$.  In practice, our code uses
Cramer's rule on the $\sigma_i\in\{\Omega,\Lambda,\Lambda'\}$ equations
of (\ref{lagrange}) to find the Lagrange multipliers:
\begin{equation}
\xi_J={-1 \over d}\left| \begin{array}{ccc}
{\partial E\over\partial \Omega} & {\partial{\cal C}\over\partial \Omega} & {\partial{\cal C'}\over\partial \Omega} \\
{\partial E\over\partial\Lambda} & {\partial{\cal C}\over\partial\Lambda} & {\partial{\cal C'}\over\partial\Lambda} \\
{\partial E\over\partial\Lambda'} & {\partial{\cal C}\over\partial\Lambda'} & {\partial{\cal C'}\over\partial\Lambda'}
         \end{array} \right|,~~~
\xi_{\cal C}={-1 \over d}\left| \begin{array}{ccc}
{\partial J\over\partial \Omega} & {\partial E\over\partial \Omega} & {\partial{\cal C'}\over\partial \Omega} \\
{\partial J\over\partial\Lambda} & {\partial E\over\partial\Lambda} & {\partial{\cal C'}\over\partial\Lambda} \\
{\partial J\over\partial\Lambda'} & {\partial E\over\partial\Lambda'} & {\partial{\cal C'}\over\partial\Lambda'}
         \end{array} \right|,~~~
\xi_{\cal C'}={-1 \over d}\left| \begin{array}{ccc}
{\partial J\over\partial \Omega} & {\partial{\cal C}\over\partial \Omega} & {\partial E\over\partial \Omega} \\
{\partial J\over\partial\Lambda} & {\partial{\cal C}\over\partial\Lambda} & {\partial E\over\partial\Lambda} \\
{\partial J\over\partial\Lambda'} & {\partial{\cal C}\over\partial\Lambda'} & {\partial E\over\partial\Lambda'}
         \end{array} \right|, \label{lagrangemultipliers}
\end{equation}
where the determinant
\begin{equation}
d=\left| \begin{array}{ccc}
{\partial J\over\partial \Omega} & {\partial{\cal C}\over\partial \Omega} & {\partial{\cal C'}\over\partial \Omega} \\
{\partial J\over\partial\Lambda} & {\partial{\cal C}\over\partial\Lambda} & {\partial{\cal C'}\over\partial\Lambda} \\
{\partial J\over\partial\Lambda'} & {\partial{\cal C}\over\partial\Lambda'} & {\partial{\cal C'}\over\partial\Lambda'}
         \end{array} \right| . \label{denom}
\end{equation}
Using equations (\ref{lagrangemultipliers}), we substitute for the
Lagrange multipliers in the remaining 7 equations of (\ref{lagrange})
to find the orbital velocity as well as the shape and structure of the
neutron stars.

The Lagrange multipliers can even be determined analytically, since
expressions for the various derivatives appearing in equations
(\ref{lagrangemultipliers}) and (\ref{denom}) can be derived from
equations (\ref{efunctional}), (\ref{jfunctional}), and
(\ref{lrs4.2.20}).  The algebra is simplified somewhat by
$\partial{\cal C}/\partial\Lambda'=0$ and $\partial{\cal
C'}/\partial\Lambda=0$.  We find $\xi_J=-\Omega$, $\xi_{\cal
C}=-\Lambda$, and $\xi_{\cal C'}=-\Lambda'$ as mathematical identities,
for any
$r,\rho_{oc},\lambda_1,\lambda_2,\rho_{oc}',\lambda_1',\lambda_2'$ and
not just along equilibrium sequences.  This provides a convenient check
of a large portion of the code, since the Lagrange variables are
determined by evaluating the determinants in equations
(\ref{lagrangemultipliers}) and (\ref{denom}) directly at every
iteration in the Newton-Raphson scheme.  Furthermore, equation
(\ref{lagrange}) then implies
\begin{equation}
dE-\Omega dJ-\Lambda d{\cal C}-\Lambda' d{\cal C'}=0,
\end{equation}
so that $dE=\Omega dJ$ along both corotating ($\Lambda=\Lambda'=0$) and
constant circulation ($d{\cal C}=d{\cal C'}=0$) sequences.
Consequently, the minima of $J$ and $E$ always coincide precisely on
our equilibrium sequences.

\section{Results}

We concentrate on binaries composed of identical stars, each with
polytropic index $n=0.5$ or $1.0$, and with a compactness parameter
$GM/Rc^2=0.125, 0.2,$ or $0.25$ in isolation.  These values of $n$ and
$GM/Rc^2$ cover an interesting region of parameter space relevant to
neutron stars.  For the AV14+UVII equation of state of Ref. \cite{WFF},
stars in the mass range $0.7 M_\odot\lesssim M\lesssim 1.7 M_\odot$
have compactness parameters $0.1\lesssim GM/Rc^2 \lesssim 0.25$ with
corresponding effective polytropic indices $1\gtrsim n\gtrsim 0.5$ (see
Table 3 of Ref. \cite{LRS3}).  For an isolated neutron star with $M=1.4
M_\odot$, the nine equations of state in Ref. \cite{ab77} which allow
this mass all yield a compactness parameter in the range $0.13\lesssim
GM/Rc^2\lesssim 0.28$.

Some physical quantities describing our neutron stars in isolation are
listed in Table \ref{isolated.spherical.models}.  These quantities can
be converted to physical units once a value of the polytropic constant
$K$ is chosen.  For instance, for $n=0.5$ and $K=1.35\times 10^{-10}$
cm$^8/$g$^2/$sec$^2$, the maximum gravitational mass $M^{max}=2.20M_\odot$,
so that $M=0.83, 1.39$, and $1.74M_\odot$ and $R=9.8, 10.3$, and $10.3$
km for $GM/Rc^2=0.125, 0.2$, and $0.25$, respectively.  For $n=1$ and
$K=1.20\times 10^5$ cm$^5/$g$/$sec$^2$, the maximum mass $M^{max}=1.600
M_\odot$ so that $M=1.15, 1.52$, and $1.599M_\odot$ and $R=13.6, 11.2$,
and $9.4$ km for $GM/Rc^2=0.125, 0.2$, and $0.25$, respectively.  Any of
our models can be scaled to a particular mass (say $1.4 M_\odot$) by an
appropriate choice of the polytropic constant $K$.

Integration of the OV equation reveals that the true maximum masses of
$n=0.5$ and $1.0$ sequences occur at log $K\rho_{oc}/c^2=-0.082$ and
$-0.497$, respectively, at which point $GM/Rc^2=0.316$ and
$GM/Rc^2=0.214$.  We note, however, that our $n=1.0$ isolated stars with
$GM/Rc^2=0.25$ are indeed stable in our PN-accurate model, since our
energy functional [eq. (\ref{isolated.energy.functional})] gives
$GM/Rc^2=0.2556$ at the maximum isolated mass (see Table
\ref{isolated.spherical.models}).

When expressing our results, the orbital angular velocity, angular
momentum, energy, and circulation are listed in terms of the
dimensionless ratios
\begin{equation}
\bar \Omega={\Omega\over (\pi G\bar \rho_o)^{1/2}},~~~
\bar J={J\over (GM_o^3R_o)^{1/2}},~~~
\bar E={E\over (GM_o^2/R_o)},~~~
\bar {\cal C}={{\cal C}\over (GM_o^3R_o)^{1/2}}, \label{dimensions}
\end{equation}
where ${\bar\rho}_o= M_o/(4\pi R_o^3/3)$.  Here $R_o$ is defined to be
the radius that an isolated polytrope, with the same $M_o$,
$K$, and $n$, would have in Newtonian theory (ie. without the $U_{PN}$
and $W_{PN}$ corrections):
\begin{equation}
R_o\equiv\xi_1 
\left[{(n+1)K\over4\pi G}\right]^{n/(3-n)}
\left({M_o\over4\pi\xi_1^2|\theta'_1|}\right)^{(1-n)/(3-n)}.\label{ro}
\end{equation}
The radius $R_o$ is a convenient unit of length because it is constant
along equilibrium sequences. 

The orbital separation $r$ is measured in harmonic coordinates while
the semi-major axes $a_i$ are measured in Schwarzschild coordinates,
which is permissible because we have two distinct small parameters in
our PN expansion, $Gm/rc^2$ and $GM/Rc^2$.  We determine when the
neutron stars touch by the PN accurate relation $r+GM_t/c^2=a_1+a_1'$,
where $M_t=2M_o+E/c^2$ is the total mass energy of the system.  Since
the radial coordinate $r$ is not a gauge invariant quantity, we choose
to make our plots as a function of the orbital angular frequency
$f=\Omega/2\pi$ ($=1/2$ the gravitational wave frequency).  The
non-dimensionalized orbital velocity $\bar \Omega$ can be converted to
a frequency $f$ in Hz via the relation
\begin{equation}
M_{1.4}f=1788\hbox{Hz}\left(5 {G M_o \over R_o c^2}\right)^{3/2}{M\over
M_o}\bar \Omega,
\end{equation}
where $M_{1.4}=M/1.4M_\odot$.
For given values of $n$ and $GM/Rc^2$,
the corresponding values of $GM_o/R_oc^2$ and $M/M_o$ can be determined
from Table \ref{isolated.spherical.models}.  For example, an $n=1$
model with $GM/Rc^2=0.2$ corresponds to $GM_o/R_oc^2=0.1467$ and
$M/M_o=0.9083$.  While $M_o$ and $M_o'$ are constant along our
equilibrium sequences, the ratio $GM/Rc^2$ does indeed vary; however,
whenever we specify a particular value of $GM/Rc^2$ we are referring to
that value at infinity (ie. for isolated neutron stars).

\subsection{Irrotational Sequences}

Figures~\ref{evsw2irrotational} and \ref{evsw2irrotational.n=1} present
irrotational (${\cal C}=0$) sequences in the equal mass case with
$n=0.5$ and $n=1$, respectively, showing the total energy $E$, relative
to its value $E_\infty$ at infinity, for $GM/Rc^2=$0.125, 0.2 and 0.25
(thick solid curves).  The energy $E_\infty$ can be calculated from
equation (\ref{isolated.energy.functional}) for each star.  Also shown
in this figure are the point-mass Newtonian equilibrium energy (thin
dashed curve), as well as the results of a Newtonian EFE treatment as
in \cite{LRS1,LRS2,LRS3,LRS4} (thick dashed line).  Figure~\ref{eosnew}
shows the dependence of $GM/Rc^2$ on the orbital frequency $f_{ISCO}$
at the ISCO for $n=0.5$ and $n=1$
irrotational sequences.  For each $n$, the curves are terminated at the
value $GM/Rc^2$ corresponding to the maximum mass.

Table \ref{irrotationalwill} presents physical quantities along
selected irrotational sequences for $n=0.5$ and $1.0$, and various
values of $GM/Rc^2$.  For each sequence, the value of $r/a_1$ marked by
a single asterisk (*) identifies the equilibrium configuration for
which $E$ (and $J$) is a minimum, ie. the ISCO.
As expected, the size of the neutron star $R/R_o$ decreases as
$GM/Rc^2$ increases.  Our $GM/R_oc^2=0$ sequences agree with the
Newtonian results of Ref. \cite{LRS3} (see their Table 1).

Tables~\ref{irrotationale} and \ref{irrotationalj} show the sizes of
the individual energy and angular momentum terms appearing in equations
(\ref{efunctional}) and (\ref{jfunctional}) for the specific sequence
$GM/Rc^2=0.2$, $n=1$.  Note that some of the relativistic corrections
are included in $E_H$ and $J_H$: for instance, at the ISCO, $E_H$ is
$1.7$ times smaller than what the Newtonian expression
[eq.(\ref{Newtonianorbitalej})] would give.  The importance of the PN
terms increases as $GM/Rc^2$ increases, and the stars behave more like
point masses.

\subsection{Corotational Sequences}

Fully relativistic calculations of binary neutron star
systems (eg. \cite{wmm,b97}) provide a nice test of our PN models.
However, such calculations are currently limited to corotating, or
nearly corotating, stars.  Figures~\ref{evsw2corotating} and
\ref{evsw2corotating.n=1} present corotating ($\Lambda=\Lambda'=0$)
sequences with $n=0.5$ and $n=1$, respectively, showing the total
energy $E$, relative to its value $E_\infty$ at infinity, for
$GM/Rc^2=0.125, 0.2,$ and 0.25 (thick solid curves).

Tables \ref{corotatingwill} presents physical quantities along selected
corotating sequences.  For each sequence, the value of $r/a_1$ marked
by a single asterisk (*) identifies the equilibrium configuration for
which $E$ (and $J$) is a minimum, the innermost {\it secularly} stable
circular orbit, while the value marked by a double asterisk (**)
identifies the innermost {\it dynamically} stable circular orbit.  The
dynamical ISCO occurs when $0=dE/dr$ at constant circulation along an
equilibrium sequence (see Ref. \cite{LRS1}).  We numerically evaluated
$dE/dr$ simply by finding equilibrium configurations at two slightly
different orbital separations and then finite differencing.  As in our
irrotational sequences, the overall size of the neutron star $R/R_o$
does indeed decrease as $GM/Rc^2$ is increased.  For corotation, the
spin of each star causes a rotational bulge such that $a_2>a_3$.  The
circulation $\cal C$ grows in magnitude as the radial separation of the
neutron stars decreases.

\subsection{Innermost Stable Circular Orbit}

Table~\ref{compare.to.lw96} presents the total angular momentum $J$,
the total mass energy $M_t=2M_o+E/c^2$, $a_2/a_1$, $a_3/a_1$ and the
orbital frequency $f$, all at the ISCO of selected irrotational
sequences.  For any given $n$, the orbital frequency $f$ can be
converted into units of Hz if either the gravitational mass $M$ or the
equation of state (ie. the polytropic constant $K$) is specified.  It
is therefore useful to express the ISCO frequency both as
$M_{1.4}f_{ISCO}$ [Hz] (when $M$ is known) and $M_{1.6}^{max}f_{ISCO}$
[Hz] (when the equation of state is known), where $M_{1.4}=M/(1.4
M_\odot)$ and $M^{max}_{1.6}=M^{max}/(1.6 M_\odot)$.  Here $M^{max}$ is
the maximum gravitational mass in isolation for a given equation of
state.  The conversion factor between $M_{1.4}f_{ISCO}$ and
$M_{1.6}^{max}f_{ISCO}$ depends on the quantity $M/M^{max}$, which can
be determined from the last two columns of Table
\ref{isolated.spherical.models}.

Table~\ref{corotating.isco} lists the orbital frequency $f_{ISCO}^{\rm
sec}$ at the innermost secularly stable orbit as well as the frequency
$f_{ISCO}^{\rm dyn}$ at the innermost dynamically stable orbit.  For
the six combinations of $n$ and $GM/Rc^2$ listed in this table, the
frequency $f_{ISCO}^{\rm dyn}$ differs by less than 4\% from the
$f_{ISCO}$ values of the corresponding irrotational sequences.  That
the spin has only a small effect on the orbital dynamics is consistent
with the small values of $T_s/|W|$ (typically on the order of 0.01 near
the ISCO) listed in Table~\ref{corotatingwill}.

\subsection{Maximum mass}

An important issue is whether the maximum mass of each neutron star
increases or decreases as the binary components are brought together.
Figure \ref{variousf} shows the rest mass $M_o$, normalized to the
maximum rest mass $M_o^{\infty, \rm max}$ at infinite separation, as a
function of the central density parameter $K \rho_{oc}^{1/n}/c^2$ for
an $n=1$ corotating binary system and at three different values of the
orbital angular frequency: $f=0$ (solid curve) corresponds to
spherical isolated stars, $M_{1.6}^{max}f=434$ Hz (long dashed curve)
corresponds to the innermost dynamically stable orbit for a star with
$M_o/M_o^{\infty, \rm max}=0.938$ ($GM/Rc^2=0.2$ at infinity), and
$M_{1.6}^{max} f=463$ Hz (short dashed curve) corresponds to the
innermost dynamically stable orbit for a star with $M_o/M_o^{\infty,
\rm max}=1$ ($GM/Rc^2=0.2556$ at infinity).  As they inspiral, stars
which begin on the stable side ($dM_o/d\rho_{oc}>0$) of the $f=0$ curve
move along horizontal lines (since $M_o$ is conserved) to the left on
this plot.

Figure \ref{mvsomega4} shows the maximum equilibrium rest mass
$M_o^{\rm max}$ as a function of the orbital angular frequency for
corotating (dashed curves) and irrotational (solid curves) sequences.
For example, the maxima of the three curves in Fig. \ref{variousf}
gives three data points for the $n=1.0$ corotating curve in Figure
\ref{mvsomega4}.  The curves are terminated at the values of $f$ for
which stars of rest mass $M_o=M_o^{\infty, \rm max}$ acquire a
dynamically unstable orbit.  We see that the maximum equilibrium mass
increases as the orbit decays, regardless of the sequence considered.
We conclude that if the maximum equilibrium mass does ever decrease in
a binary system, it can do so only through PN tidal terms, or higher
order PN terms, which we are neglecting in this paper (see
\S\ref{future.work}).  Note that we have not determined from our
equilibrium analysis, which of our equilibria are stable.  The
exception is the $f=0$ (infinite separation) sequence, for which the
turning point on the $M$ versus $\rho_{oc}$ plot marks the onset of
radial instability to collapse.

\section{Discussion}
\subsection{Summary of Results}

We have extended the work of previous studies in a number of ways.
We have improved upon purely Newtonian EFE treatments by including PN
effects, and improved upon point-mass PN treatments by including both
Newtonian and PN finite-size effects.  The main message of Appendix B
is that equation (\ref{isolated.energy.functional}) is the
PN accurate energy functional of an isolated non-rotating star of
arbitrary polytropic index $n$.
Nowhere in the derivation of equation
(\ref{isolated.energy.functional}) is equilibrium assumed,
but in equilibrium the trial density function we implement agrees to PN
order with the density profile obtained by integrating the OV
equation.  The PN coefficients $l_1$ and $l_2$ appearing in equation
(\ref{isolated.energy.functional}) are listed for various polytropic
indices $n$ in Table \ref{structureconstants}.
Figure~\ref{isolated.spherical} demonstrates that using equation
(\ref{isolated.energy.functional}) yields reliable results (compare the
long dashed curve to the OV results given by the solid curve),
especially for soft equations of state (which have a maximum mass at
small $\rho_{oc}$).  One particularly nice feature of our energy
functional is that it correctly exhibits a maximum mass even for stiff
equations of state.  Various physical quantities for our isolated
spherical models are listed in Table~\ref{isolated.spherical.models}.

Figures \ref{evsw2irrotational} and \ref{evsw2irrotational.n=1},
present the equilibrium energy $E$ as a function of the orbital
frequency $f$ along irrotational sequences with various values of the
polytropic index $n$ and compactness parameters $GM/Rc^2$.  Note that
for a given $M$ and $GM/Rc^2$, $n=1$ polytropes are more centrally
concentrated than $n=0.5$ polytropes, so that the former behave more
like point masses (ie. Newtonian effects which destabilize the orbit
are less important for $n=1$) and have a slightly larger
$M_{1.4}f_{ISCO}$.  This same tendency can be seen by comparing the
$n=0.5$ and $n=1.0$ corotating sequences presented in Figures
\ref{evsw2corotating} and \ref{evsw2corotating.n=1}, respectively.  In
these plots the minimum of the equilibrium energy $E$ marks the onset
of an orbital secular instability.  The dynamical instability sets in
at higher orbital frequency, as can be seen from Table
\ref{corotatingwill} or \ref{corotating.isco}.

Tables \ref{irrotationalwill} and \ref{corotatingwill} present various
quantities along selected sequences for comparisons with future
studies.  The Tables \ref{irrotationale} and \ref{irrotationalj}
present the relatives sizes of the terms which make up our energy and
angular momentum functionals; these tables allow us to explicitly
examine the relative importance of the various effects which determine
the orbital dynamics and the interior structure.  Comparing the
magnitudes of $W_t$ and $E_{OI}$, it is clear that both tidal and
relativistic terms play an important role.   Tables
\ref{compare.to.lw96} and \ref{corotating.isco} summarize some of the
important quantities at the ISCO of our irrotational and corotating
sequences, respectively.  Note that for fixed $n$, $M_{1.4}f_{ISCO}$
always increases with $GM/Rc^2$, since more compact stars behave more
like point masses.

Although real neutron star sequences are probably close to
irrotational, fully relativistic calculations of binary systems are
currently limited to corotating sequences.  By looking for turning
points in the equilibrium energy, it is straightforward to determine
the secular ISCO frequency $f^{sec}_{ISCO}$ in such corotating
sequences.  It is often assumed that the actual ISCO frequency
$f_{ISCO}$ of an irrotational sequence does not differ drastically from
the frequency $f^{sec}_{ISCO}$ determined from corotating
calculations.  Our results allow us to quantify this difference.  For
$n=0.5$ we find that the irrotational $f_{ISCO}$ is approximately 17\%
larger than the corresponding $f^{sec}_{ISCO}$, while for $n=1$ the
difference is approximately 20\% (compare entries in
Tables~\ref{compare.to.lw96} and \ref{corotating.isco}).  For any
polytropic index $n$, this difference depends only very weakly on
$GM/Rc^2$.

If multiple binary neutron star coalescence events are observed by
instruments such as LIGO, then tables like Table~\ref{compare.to.lw96},
or figures like Figure~\ref{eosnew}, will give information regarding
the equation of state.  For instance, consider a simple world in which
binary components always had identical masses, although allow for
varying neutron star mass from one binary to the next.  Suppose further
that all neutron stars obeyed the same (but unknown) polytropic
equation of state.  A single coalescence observation would yield a
value of $M_{1.4}f_{ISCO}$, since the gravitational mass $M$ is encoded
in the early inspiral waveform and since the ISCO frequency $f_{ISCO}$
can be determined by the onset of the orbital plunge.  However, the
equation of state and compactness parameter $GM/Rc^2$ could not be
uniquely determined from a single observation, since we do not yet know
which curve (if either) in Figure~\ref{eosnew} applies.  For each value
of $n$ being considered, the corresponding maximum gravitational mass
$M^{max}$ can be immediately determined from the observed value of
$f_{ISCO}$ and from the $M^{max}_{1.6}f_{ISCO}$ column in
Table~\ref{compare.to.lw96}.  An additional observation of a
coalescence event with a different $f_{ISCO}$ would yield another value
of $M^{max}$ for each $n$.  The polytropic index which consistently
gives the same $M^{max}$ regardless of the observed $f_{ISCO}$ must
then represent the true equation of state.  Tables similar to
Table~\ref{compare.to.lw96} will therefore be helpful in testing
realistic equations of state once coalescence observations are
available.

There is only one equation of state for neutron stars, even if
knowledge of it is still uncertain.  For a given equation of state, one
might ask how $f_{ISCO}$ depends on the mass of the binary components.
For polytropes, fixing $n$ and $K$ also fixes the maximum gravitational
mass $M^{max}$ in isolation.  From the $M_{1.6}^{max}f_{ISCO}$ column
of Table~\ref{compare.to.lw96}, we see that for $n=0.5$ the general
trend of $f_{ISCO}$ is for it to decrease as the stellar mass $M$ (or
$GM/Rc^2$) increases, while for $n=1$ the frequency $f_{ISCO}$
increases with $M$.  We therefore conclude that the relation between
$f_{ISCO}$ and $M$ depends strongly on the equation of state.  The
reason for this dependence can be traced to two competing effects.
Namely, as we consider more compact (larger $M$) stars, PN orbital
destabilization effects become stronger, while Newtonian
destabilization effects become weaker.

For the cases listed in Table~\ref{compare.to.lw96}, the total angular
momentum $J$ is $0.96$ to $1.10$ times the total energy squared
$M_t^2=(2M_o+E)^2$ (in gravitational units, $G=c=1$) at the ISCO.  In
those cases with $J>M_t^2$, angular momentum would need to be either
radiated away or carried away by ejected matter before a black hole can
form (see Ref. \cite{eh96} for a discussion).  We find $J>M_t^2$ for
all $n=0.5$ sequences with $GM/Rc^2\lesssim 0.21$, while for $n=1$ this
criterion is satisfied when $GM/Rc^2\lesssim 0.18$.

Figure \ref{variousf} presents the equilibrium rest mass $M_o$ as
function of the central density for various orbital frequencies $f$ for
an $n=1$ corotating sequence.  We see that, for the terms we have
included in our energy functional, the maximum mass in a close binary
is slightly larger than in isolation.  Figure \ref{mvsomega4}
summarizes our maximum mass results for both irrotational and corotating
sequences, and for $n=0.5$ and $1.0$.  The irrotational sequences
maintain a lower maximum mass than their corresponding corotating
sequences, because corotating stars (a) have more rotational support
and (b) are more ellipsoidal so that the Newtonian tidal field has more
of a stabilizing effect.  The relative increases in $M_o^{max}$ is more
gradual for $n=0.5$ than $n=1.0$, because $n=0.5$ stars near the
maximum mass are much more centrally condensed than $n=1.0$ stars near
their maximum mass: from Table \ref{isolated.spherical.models} we see
that both the central density parameter $K\rho_{oc}/c^2$ and the
compactness parameter $GM/Rc^2$ are much larger for the maximum mass
$n=0.5$ star (log $q_c=0.12$, $GM/Rc^2=0.39$) than the $n=1$ star (log
$q_c=-0.39$, $GM/Rc^2=0.26$).  Consequently, the $n=0.5$ stars near the
maximum mass have smaller moments of inertia and behave more like point
masses, so that stabilizing hydrodynamic effects play a less
significant role.  Our results suggest that the maximum equilibrium
mass increases as the orbit decays, regardless of the sequence
considered.  However, if the maximum equilibrium mass does decrease in
a binary system, it must be a result of the PN tidal terms, or of
higher order PN corrections, that we are neglecting (see
\S\ref{future.work}).

\subsection{Comparison with Other Work}

Recently, Taniguchi and Nakamura \cite{tn96} have applied the
ellipsoidal approximation to a neutron star-black hole binary.  They
also compute irrotational and corotating sequences, but only for $n=0$
polytropes.  In place of the hybrid-PN orbital energy contribution
$E_{H}$, they use a generalized pseudo-Newtonian potential constructed
to fit the ISCOs of the hybrid P$^2$N treatment of Kidder, Will, and
Wiseman \cite{kww}.  We improve upon the work of Taniguchi and Nakamura
by including the relativistic corrections $U_{PN}$ and $W_{PN}$ to the
internal energy of the neutron star, and by calculating sequences of
fixed $M_o/R_o$, instead of fixed $M/R$ which varies as two neutron
stars inspiral.  Despite these differences our results do agree
qualitatively: for a given equation of state more compact stars have
larger ISCO frequencies, and for a fixed compactness parameter the ISCO
along irrotational sequences is larger than the secular ISCO along
corotating sequences.

Taniguchi and Shibata \cite{ts97} have recently presented PN EFE
analyses of binary neutron star systems which is rather similar to our
own.  The main differences are that (1) they derive equilibrium
conditions from a tensor virial method, whereas we use an energy
variational method, (2) they include all terms, up to quadrupole order,
which are formally PN order, while we have neglected PN tidal terms and
PN corrections to the internal fluid motion, (3) they consider only
corotating sequences, while we also treat the more realistic
irrotational sequences, and (4) they implement a simple Lane-Emden
trial density function, as opposed to our PN-accurate density function
[see eq. \ref{ov-based}].  Our treatment does somewhat better than
Taniguchi and Shibata's at matching the numerical, PN-accurate results
of Shibata \cite{shibata}.  For instance, for $n=0.5$ and
$GM_o/R_oc^2=0.02$, Shibata finds secular ISCO values of $\bar E$,
$\bar J$, and $\bar \Omega$ equal to
$-1.236$, $1.457$, and $0.31$, respectively; compare this to our values
$-1.246$, $1.458$, and $0.316$, and to Shibata and Taniguchi's values
$-1.269$, $1.443$, and $0.327$.  
We find a similar level of agreement with Shibata along other
sequences:  for $n=1.0$ and $GM_o/R_oc^2=1/60$, Shibata finds secular
ISCO values of $\bar E$, $\bar J$, and $\bar \Omega$ equal to
$-1.146$, $1.386$, and $0.35$, while our treatment gives
$-1.155$, $1.384$, and $0.364$.

In order to examine the ISCO in irrotational sequences, Lai and Wiseman
\cite{lw96} have combined the Newtonian ellipsoidal equations of motion
\cite{ls95} with the hybrid P$^2$N point mass equations of motion of
Kidder, Will, and Wiseman.  This treatment does not include the PN
coupling between the orbital motion and internal structure ($E_{OI}$
and $J_{OI}$ in this paper), nor does it include PN corrections to the
internal and self-gravitational energy ($U_{PN}$ and $W_{PN}$ in this
paper).  Nevertheless, our results are in excellent agreement with
those of Lai and Wiseman (see the comparison in
Table~\ref{compare.to.lw96}):  for the three values of $GM/R_oc^2$ for
which a direct comparison is possible, we find agreement to better than
4\% in $f_{ISCO}$ and to better than about 2\% in the ratios $a_2/a_1$
and $a_3/a_1$.  The comparisons of $a_2/a_1$ and of $a_3/a_1$ should be
viewed with some caution because of the freedom to measure distances
with various coordinate systems in general relativity.  In the
Newtonian limit our results agree precisely with the Newtonian results
of Lai and Wiseman [see their equations (8) and (9)], as they must
since both treatments are extensions of the same Newtonian ellipsoidal
approximation.

In the point mass limit, the results of Lai and Wiseman are accurate to
P$^2$N order, whereas our results are accurate only to first PN order.  We
have not implemented the point mass P$^2$N orbital energy and angular
momentum corrections in our energy functional, because doing so causes
$dE=\Omega\,dJ$ to hold to P$^2$N order but not exactly along our
sequences, and because our functional would be neglecting other P$^2$N
terms which are larger.  However, we have explicitly tested the
importance of the point mass P$^2$N corrections by adding them to our
energy and angular momentum functions.  We find that for
$GM/Rc^2=0.25$, the ISCO frequency for our $n=0.5$ and $n=1$
irrotational sequences is decreased by less than 4\%, while for
$GM/Rc^2=0.125$ and $0.2$ this frequency is increased by about 2\% or
less.

Baumgarte et al. \cite{b97} have recently calculated in full general
relativity quasi-equilibrium corotating sequences of binary neutron
stars obeying a polytropic equation of state.
Their computations are the most reliable to date for the neutron star
structure, maximum equilibrium mass, and determination of the secular
ISCO.  Over the range of orbital separations they consider, they find
the maximum equilibrium mass to increase slightly as the orbit decays,
in agreement with the results presented in Figures~\ref{variousf} and
\ref{mvsomega4}.  This agreement is reassuring, since these
relativistic calculations essentially represent the ``true'' solution.
Figure \ref{compare.with.tom} compares our binding energy results with
those of Ref. \cite{b97} for an $n=1$ corotating sequence.

The maximum equilibrium rest mass results presented in
Figures~\ref{variousf} and \ref{mvsomega4} are also in qualitative
agreement with the results of Lai \cite{l96}: the maximum mass of a
neutron star seems to increase as the orbit decays.  The main
differences between our energy functional and that of Lai are that (a)
Lai's adopted equation of state is based on a degenerate,
non-relativistic Fermi gas of nucleons ($n=1.5$ polytrope), while we
have concentrated on $n=0.5$ and $n=1$ polytropes, (b) Lai has included
a correction $\Delta E_{int}$ as in Ref. \cite{st83} accounting for the
tendency of neutrons to become mildly relativistic, and (c) we use
equation (\ref{isolated.energy.functional}) for the self energy of a
polytrope, whereas Lai uses equation (\ref{st.eeq}).  Our maximum mass
results are also completely consistent with the recent analytic
treatments of Brady and Hughes \cite{bh97} and of Wiseman \cite{w97},
which show that there can be no decrease in the maximum equilibrium
rest mass at first PN order, presuming tidal effects can be neglected.
However, we cannot be sure that PN tidal effects would not change our
results (see \S\ref{future.work}).  Nevertheless, our results are
consistent with the relativistic integrations of Baumgarte et al.
(1997), which automatically include all PN tidal terms.  This agreement
provides strong support of our analysis, at least in the case of
corotating binaries.

\subsection{Directions for Future Work} \label{future.work}

Recently, Wilson, Mathews, and Marronetti \cite{wmm} have presented
numerical simulations which suggest that otherwise stable neutron stars
may individually collapse to black holes when brought into a close
binary orbit.  Our results suggest that the maximum equilibrium mass
increases as the orbit decays.  In this section, however, we discuss
two important caveats which leave open the possibility that the stars
could collapse prior to the orbital plunge: (1) the PN tidal terms
neglected in this paper may affect the maximum equilibrium mass, and
(2) even if the maximum equilibrium mass does increase in binaries,
some of the allowable equilibria may be {\it unstable} to collapse.

The largest uncertainty in our results may come from neglected PN tidal
terms.  Chandrasekhar \cite{c65} and Chandrasekhar and Nutku
\cite{cn69} have derived the PN corrections to the energy and angular
momentum for arbitrary fluid configurations, which could be used to
determine the PN corrections for our binary star system.  The
tidal components of these corrections are absent from our present
analysis.  For example, consider the PN correction from Ref.
\cite{cn69}'s equation (68) which is proportional to $\int
\rho\Phi^2\,d^3x$, where $\Phi$ is the usual Newtonian gravitational
potential.  Inside a star, the potential $\Phi$ can be decomposed into
an internal contribution $\Phi_{int}$ due to that star, and an external
contribution $\Phi_{ext}$ due to the orbital companion.  The integral
$\int \rho\Phi_{int}^2\,d^3x$ then contributes to the
self-gravitational PN energy $W_{PN}$, while the cross term
$\int\rho\Phi_{ext}\Phi_{int}\,d^3x$ gives a coupling as in $E_{OI}$.
The integral $\int \rho\Phi_{ext}^2\,d^3x$ to lowest order helps give
the point mass PN orbital energy like in Ref. \cite{kww}, and for
finite-size stars this integral can be evaluated in a multipole
expansion with the higher order terms accounting for PN tidal effects.
A careful investigation of PN tidal effects seems
worthwhile.

%
%
%
%
%

We also note that we have not tested our equilibrium configurations for
stability against collapse to black holes: even if the maximum
equilibrium mass does increase as the orbit decays, this does not
necessarily mean the maximum stable mass follows the same trend.  The
EFE energy variational approach provides a straightforward means of
testing the stability of equilibrium models, which could be exploited
in future studies.  In particular, the second order variation of the
energy functional can be used to identify any instability to collapse.

\acknowledgments

We would like to thank Phil Arras for a careful checking of the
Appendices, as well as Thomas Baumgarte, \'{E}anna Flanagan, Larry
Kidder, Dong Lai, Saul Teukolsky, and Alan Wiseman for useful
discussions.  This work was supported by NSF grant AST 91-19475 and
NASA grant NAG5-2809 at Cornell, and by NSF grant AST 96-18524 and NASA
grant NAG5-3420 at Illinois.  FAR is supported by an Alfred P. Sloan
Foundation Fellowship.


\appendix
\section{Post-Newtonian Expansion of the OV Equation} \label{PN.expand.OV}

In this appendix we expand the OV equation (see, eg. \cite{st83}) to PN
order and thereby derive the first PN corrections to the Newtonian
Lane-Emden functions for an isolated spherical polytrope in
equilibrium.  Our results are useful for generating trial density
functions in \S\ref{ov.based.density.function} accurate to PN order.

Let $\rho_o(r)$ be the rest mass density of a spherically symmetric
distribution of matter at the Schwarzschild radius $r$.  Let $m(r)$ be
the enclosed total mass-energy, and define the quantity $\xi$ and the
non-dimensional functions $\vartheta(\xi)$ and $\nu(\xi)$ by
\begin{equation}
\rho_o=\rho_{oc}\vartheta^n,~~~r=a\xi,~~~m=-4\pi a^3\rho_{oc}\nu\xi^2,
\label{non.dimensional}
\end{equation}
where $\rho_{oc}=\rho_o(0)$ is the central rest mass density, so that
$\vartheta(0)=1$.  In the Newtonian limit, the functions
$\vartheta(\xi)$ and $\nu(\xi)$ reduce to the usual Lane-Emden
functions $\theta$ and $\theta'$.  Here $a=R/\xi_I$ is a scale factor,
with $R$ being the radius of the star in Schwarzschild coordinates and
$\xi_I$ being defined by the condition $\vartheta(\xi_I)=0$.  For a
polytropic equation of state
\begin{equation}
P=K\rho_o^{1+{1\over n}}
\end{equation}
the OV equations yield a hydrostatic equilibrium solution if (see, eg.,
Ref. \cite{t65})
\begin{eqnarray}
{d\vartheta \over d \xi} & = & {
\nu\left(1+(n+1)q_c\vartheta\right)\left(1-q_c\vartheta^{n+1}\xi/
\nu\right) \over \left(1+2(n+1)q_c\nu\xi\right)}, \label{ov1}\\
{1 \over \xi^2}{d(\nu \xi^2)\over d\xi} & = &
-\vartheta^n\left(1+nq_c\vartheta\right), \label{ov2} \\
{K \rho_{oc}^{1/n}\over c^2} & = & q_c, \label{equilibrium.rho}
\end{eqnarray}
where we have defined
\begin{equation}
q_c\equiv{4\pi\rho_{oc}G a^2 \over (n+1)c^2}. \label{qc}
\end{equation}
Evaluating the last equation of (\ref{non.dimensional}) at the surface,
we see that for an equilibrium configuration the scale factor $a$
satisfies
\begin{equation}
4\pi a^3\rho_{oc}={M \over \xi_I^2 |\nu_I|}, \label{scale.factor}
\end{equation}
where $M$ is the total mass-energy and $\nu_I=\nu(\xi_I)$.  In the
Newtonian limit ($c\rightarrow\infty$), $q_c\rightarrow 0$,
$\vartheta\rightarrow\theta$ and $\nu\rightarrow\theta'$, where
$\theta$ is the usual Lane-Emden function.  If we define the functions
$\alpha$ and $\beta$ as the PN components in the expansions
\begin{eqnarray}
\vartheta & = & \theta+q_c\alpha+{\cal O}(q_c^2), \label{alpha} \\
\nu & = & \theta'+q_c\beta+{\cal O}(q_c^2), \label{beta}
\end{eqnarray}
then equations (\ref{ov1}) and (\ref{ov2}) become
\begin{eqnarray}
{d\alpha \over d\xi} & = &
\beta+(n+1)\theta\theta'-\theta^{n+1}\xi-2(n+1)(\theta')^2\xi,
\label{dalpha}\\
{1 \over \xi^2}{d(\beta \xi^2)\over d\xi} & = &
-n(\alpha\theta^{n-1}+\theta^{n+1}). \label{dbeta}
\end{eqnarray}
In deriving equations (\ref{dalpha}) and (\ref{dbeta}) we have used the
Lane-Emden equation to cancel all terms of Newtonian order.  We note
that the functions $\alpha$ and $\beta$ depend only on the polytropic
index $n$ and can be numerically integrated simultaneously with the
Lane-Emden equation.  The numerical integrations can be started
slightly away from $\xi=0$ with the help of the approximations, for
$\xi<<1$:
\begin{eqnarray}
\alpha & \approx & -{\xi^2\over 3}(n+2), \\
\beta  & \approx & -{n \over 3} \xi.
\end{eqnarray}

The surface occurs when
$0=\vartheta(\xi_I)=\theta(\xi_I)+q_c\alpha(\xi_I)=(\xi_I-\xi_1)\theta_1'+q_c\alpha(\xi_1)$,
to PN order, where we have Taylor expanded and used $\theta(\xi_1)=0$
to obtain the last equality.  Therefore,
\begin{equation}
\xi_I=\xi_1-q_c{\alpha_1\over\theta_1'}+{\cal O}(q_c^2), \label{surface}
\end{equation}
where $\alpha_1=\alpha(\xi_1)$.  Similarly, Taylor expanding equation
(\ref{beta}) about the surface at $\xi=\xi_I$ gives
\begin{equation}
\nu_I=\theta_1'+q_c{2\alpha_1\over\xi_1}+q_c\beta_1+{\cal O}(q_c^2),
\end{equation}
where $\nu_I=\nu(\xi_I)$ and $\beta_1=\beta(\xi_1)$.

In the special case $n=1$ it is well known that the solution to the
Lane-Emden equation is $\theta=\sin(\xi)/\xi$.  Equations
(\ref{dalpha}) and (\ref{dbeta}) can also be solved analytically when
$n=1$:
\begin{eqnarray}
\alpha & = & \theta^2-\theta-2\xi\theta'\theta-3\xi^2\theta^4-
3\theta\int_0^\xi\xi\theta^3\,d\xi+
9\left(\theta'+{\theta\over\xi}\right)\int_0^\xi\xi^2\theta^3\,d\xi,
\label{alpha.n=1} \\
\beta & = & 3\xi\theta^2-\theta'+2\theta\theta'
-3\xi^2\theta^3\theta'+2\xi\theta'^2
-3\theta'\int_0^\xi\xi\theta^3\,d\xi
-{9(\theta+\xi^2\theta+\xi\theta')\over\xi^2}\int_0^\xi\xi^2\theta^3\,d\xi.
\end{eqnarray}
These equations provide a convenient check of our numerical
integrations in Appendix B.

\section{Post-Newtonian Self-Energy Terms}\label{derive.l1.and.l2}

We now derive the PN self energy terms of a spherically symmetric fluid
which is instantaneously at rest.  Our approach generalizes the method
of Ref. \cite{zn71}.  Let us introduce the enclosed rest mass $m'$ and
the proper radial coordinate $r'$, which are related to the enclosed
total mass-energy $m$ and the Schwarzschild radius $r$ by
\begin{equation}
dm'=\rho_o4\pi r'^2\,dr'=\left(1-{2Gm\over
rc^2}\right)^{-1/2}\rho_o4\pi r^2\,dr=\left(1-{2Gm\over
rc^2}\right)^{-1/2}\left(1+{u\over c^2}\right)^{-1}dm,
\label{differentials}
\end{equation}
where $u$ is the internal energy density.  To PN order, the energy of
such a system, excluding rest mass energy, is [see equations (6.9.3),
(6.9.8), (6.9.10) and (6.9.18) of Ref. \cite{st83}]

\begin{eqnarray}
E & = & c^2\int_0^Mdm - c^2\int_0^{M_o}dm' \label{exact.energy} \\
  & = & \int_0^{M_o}u\,dm'-G\int_0^{M_o}{m' \over
  r'}dm'+I_1+I_2+I_3+I_4+I_5, \label{energy.functional}
\end{eqnarray}
where $M_o$ is the total rest mass, and
\begin{eqnarray}
I_1 & = & -{G \over c^2}\int_0^M u{m\over r}dm, \\
I_2 & = & -{G^2 \over 2c^2}\int_0^M \left({m\over r}\right)^2dm, \\
I_3 & = & -{G \over c^2}\int_0^M{dm \over r}\int_0^m u\,dm, \\
I_4 & = & {G^2 \over c^2}\int_0^M{dm \over r}\int_0^m {m\over r}\,dm, \\
I_5 & = & -{G^2 \over c^2}\int_0^M{m\,dm \over r^4}\int_0^r mr\,dr.
\end{eqnarray}
Since we are working only to PN order, the quantities $m$ and $r$ can
be replaced with $m'$ and $r'$ in $I_1$ to $I_5$, if desired.

Note that equations (\ref{exact.energy}) and (\ref{energy.functional})
do {\it not} require that the fluid be in equilibrium: the density
profile $\rho_o$ and internal energy profile $u$ need to be spherically
symmetric, but are otherwise arbitrary.  Let us now restrict our
attention to fluids obeying a polytropic equation of state
\begin{equation}
P=K\rho_o^{1+{1\over n}},~~~u=nK\rho_o^{1/n}, \label{eos}
\end{equation}
where $K$ is a constant, but allow the density profile $\rho_o$ to
remain arbitrary for the moment.  We wish to find the contribution to
the energy beyond that given by the usual Newtonian expressions for the
internal energy $U$ and the self-gravitational energy $W$ [see
equations (\ref{lrs4.2.5}) and (\ref{lrs4.2.6})], ie.
\begin{equation}
\Delta E\equiv
E-\left(k_1K\rho_{oc}^{1/n}M_o-k_2G\rho_{oc}^{1/3}M_o^{5/3}\right),
\label{define.epn}
\end{equation}
where it is understood that all terms of order P$^2$N and higher are
dropped.  Inspection of equations (\ref{energy.functional}),
(\ref{eos}) and (\ref{define.epn}) reveals that $\Delta E$
is composed of two types of terms: those which depend on $K$ and those
which do not; we write
\begin{equation}
\Delta E=U_{PN}+W_{PN}, \label{deltaeself}
\end{equation}
where we have defined
\begin{eqnarray}
U_{PN} & = & \int_0^{M_o}u\,dm'-k_1K\rho_{oc}^{1/n}M_o+I_1+I_3,
\label{upn.app} \\
W_{PN} & = & -G\int_0^{M_o}{m' \over
r'}dm'+k_2G\rho_{oc}^{1/3}M_o^{5/3}+I_2+I_4+I_5. \label{wpn.app}
\end{eqnarray}
The quantity $U+W+U_{PN}+W_{PN}$ is the energy of a spherical symmetric
configuration with an {\it arbitrary} density profile, to PN order.

If we restrict our attention to a set of well-chosen density functions,
then the integrals in equations (\ref{upn.app}) and (\ref{wpn.app}) can
be evaluated.  In particular, if we consider a family of density
profiles of the form $\rho_o(r')\propto f(r')$ [or equivalently
$\rho_o(r)\propto g(r)$], where $f(r')$ [or $g(r)$] is arbitrary, then
\begin{eqnarray}
U_{PN} & = & -l_1{G \over c^2}K \rho_{oc}^{{1\over n}+{1\over
3}}M_o^{5/3}, \label{l1} \\
W_{PN} & = & -l_2{G^2 \over c^2}\rho_{oc}^{2/3}M_o^{7/3}. \label{l2}
\end{eqnarray}
Here the dimensionless coefficients $l_1$ and $l_2$ are determined by
evaluating the integrals in equations (\ref{upn.app}) and
(\ref{wpn.app}) for the chosen family of density profiles.

\subsection{Lane-Emden Trial Density Functions}

As a concrete example, consider density profiles which are related to
the non-dimensional Lane-Emden function $\theta$ by
\begin{equation}
\rho_o^{\rm LE}(r')=\rho_{oc} \theta^n,~~~r'={R'\over \xi_1}\xi,
\label{lane-emden}
\end{equation}
where $R'$ is the proper radius of the star and the central density
$\rho_{oc}$ does {\it not} necessarily have the value which would allow
for equilibrium at Newtonian order.  The superscript ``LE'' appearing
in equation (\ref{lane-emden}) is a reminder that we are limiting
ourselves to a single family of trial density profiles based upon
Lane-Emden functions.  Since density profiles of this form differ at
least at PN order from the OV solution which minimizes the energy given
by equation (\ref{exact.energy}), there is {\it no} value of
$\rho_{oc}$ for which the fluid is in equilibrium to PN
order.\footnotemark \footnotetext{The exception to this statement
occurs for $n=0$.  In this case, both the purely Newtonian and the
exact general relativistic density profile are of the form
$\rho_o=$const.}  Using equation (\ref{lane-emden}) in equations
(\ref{upn.app}) and (\ref{wpn.app}) we obtain simply
\begin{eqnarray}
l_1^{\rm LE} & = & {I_1+I_3 \over -{G \over c^2}K \rho_{oc}^{{1\over
n}+{1\over 3}}M_o^{5/3}}, \label{l1le} \\
l_2^{\rm LE} & = & {I_2+I_4+I_5 \over -{G^2 \over
c^2}\rho_{oc}^{2/3}M_o^{7/3}}. \label{l2le}
\end{eqnarray}
From here it is straightforward to write $l_1^{\rm LE}$ and $l_2^{\rm
LE}$ as integrals over Lane-Emden functions, and these coefficients
therefore depend only on the polytropic index $n$.  Equations
(\ref{l1le}) and (\ref{l2le}) define the coefficients used to produce
the short dashed curves in Figure~\ref{isolated.spherical}.

In the following subsection we go through a considerable amount of work
to derive $l_1$ and $l_2$ for more accurate trial density functions.
However, the coefficients given by equations (\ref{l1le}) and
(\ref{l2le}) already represent a significant improvement over that used
previously in the literature (compare the short-dashed and dot-dashed
curves in Figure \ref{isolated.spherical}).  The reason for this
improvement is that we have kept both of the PN terms [eqs. (\ref{l1})
and (\ref{l2})] which result naturally from the subtraction in equation
(\ref{define.epn}).  Previous analyses have used equilibrium relations
to eliminate $K$ in equation (\ref{l1}) and forced $U_{PN}$ to scale
like $W_{PN}$, resulting in an energy functional which is valid only
along equilibrium sequences and therefore should not be used in an
energy variational treatment [see the discussion surrounding eq.
(\ref{st.eeq}) below].  Nowhere in the derivation of equations
(\ref{l1le}) and (\ref{l2le}) have we assumed the star is in
equilibrium.

\subsection{OV-Based Trial Density Functions}\label{ov.based.density.function}

The disadvantage of being restricted to Lane-Emden trial density
functions is that the energy minimum of equation
(\ref{energy.functional}) then differs at PN order from the actual
energy minimum.  We know that fluid configurations which obey the OV
equation are precisely those necessary to minimize the energy given by
equation (\ref{exact.energy}).  This suggests that we construct a
family of trial density functions based upon the solution to the OV
equation, namely
\begin{equation}
\rho_o(r)=\rho_{oc}(\theta+q_c\alpha)^n,~~~r={R \over \xi_I}\xi,
\label{ov-based}
\end{equation}
where $R$ is the stellar radius in Schwarzschild coordinates, the
central density $\rho_{oc}$ is {\it not} necessarily the value required
by the OV equation for equilibrium [ie. eqs.  (\ref{equilibrium.rho})
and (\ref{scale.factor}) need not be satisfied],  and
the function $\alpha$ is obtained by solving equations (\ref{dalpha})
and (\ref{dbeta}).  The expansion
parameter $q_c$ is defined by equation (\ref{qc}) with $a=R/\xi_I$, ie.
\begin{equation}
q_c={4\pi\rho_{oc}G R^2 \over \xi_I^2(n+1)c^2}. \label{qc.ov-based}
\end{equation}
An energy variational method based upon this family of trial density
functions can identify the minimum of equation
(\ref{energy.functional}) to PN order.  There are an infinite number of
other families which satisfy this criterion [for instance, we could
generate one other such family by replacing $\rho_{oc}$ with its
equilibrium value in eq. (\ref{qc.ov-based})].  However, the trial
density functions defined by equations (\ref{ov-based}) and
(\ref{qc.ov-based}) are independent of the polytropic constant $K$ and
hence make it easy to track which terms do and do not depend on $K$.
We note that the last equation of (\ref{non.dimensional}) does not
apply here, although by construction it is satisfied to PN order when
$\rho_{oc}$ has its equilibrium value.  [To derive the enclosed total
mass-energy even for non-equilibrium configurations, use $m=\int\rho
4\pi r^2\,dr$ with $\rho=\rho_o(1+u/c^2)$.]

Our plan is now to evaluate $l_1$ and $l_2$ as defined by equations
(\ref{l1}) and (\ref{l2}), using the family of trial density functions
given in equation (\ref{ov-based}).  We begin with the first integral
appearing in equation (\ref{upn.app}) and work to PN order:
\begin{eqnarray}
\int_0^{M_o}u\,dm' & = & \int_0^M u \left(1-{2Gm\over r
c^2}\right)^{-1/2}\left(1+{u\over c^2}\right)^{-1}\,dm \\
& = & \int_0^M u\left(1+{Gm\over r c^2}-{u\over c^2}\right)\,dm \\
& = & \int_0^M u\left(1-{u\over c^2}\right)\,dm-I_1, \label{udmprime}
\end{eqnarray}
where we have used equation (\ref{differentials}) and expanded.
We now focus our attention on the integral
\begin{eqnarray}
\int_0^M u\left(1-{u\over c^2}\right)\,dm & = & 4 \pi n \int_0^R P
r^2\,dr \\
& = & 
nK\rho_{oc}^{1/n}{4\pi R^3\rho_{oc}\over \xi_I^3}
\int_0^{\xi_I}(\theta+q_c\alpha)^{n+1}\xi^2\,d\xi \\
& = & 
nK\rho_{oc}^{1/n}{4\pi R^3\rho_{oc}\over \xi_I^3}
\int_0^{\xi_I}\theta^{n+1}\left(1+(n+1){q_c\alpha\over
\theta}\right)\xi^2\,d\xi \\
& = & 
nK\rho_{oc}^{1/n}{4\pi R^3\rho_{oc}\over \xi_I^3}
\left(\int_0^{\xi_I}\theta^{n+1}\xi^2\,d\xi
+(n+1)q_c\int_0^{\xi_I}\alpha\theta^n\xi^2\,d\xi\right). \label{udm}
\end{eqnarray}
Note that since we are working
only to PN order, it is justified to end all the integrals in this
appendix at $\xi_1$ [where $\xi_1$ is defined by
$\theta(\xi_1)=0$].  For instance, since $\theta\lesssim q_c$
in the range $\xi_I<\xi<\xi_1$, and since $\xi_1-\xi_I\sim q_c$,
\begin{equation}
\int_0^{\xi_I}\theta^{n+1}\xi^2\,d\xi=\int_0^{\xi_1}\theta^{n+1}\xi^2\,d\xi+{\cal
O}(q_c^{n+2})={k_1\over n}\xi_1^2|\theta_1'|+{\cal O}(q_c^{n+2}).
\label{justified}
\end{equation}
Therefore equation (\ref{udm}) becomes
\begin{equation}
\int_0^M u\,dm = 
k_1K\rho_{oc}^{1/n}\xi_1^2|\theta_1'|{4\pi R^3\rho_{oc}\over \xi_I^3}
\left(1+{n(n+1)q_c\over
k_1\xi_1^2|\theta_1'|}\int_0^{\xi_1}\alpha\theta^n\xi^2\,d\xi\right)
\label{final.udm}
\end{equation}

In order to write our final answer in the form of equation
(\ref{deltaeself}), we need to derive a relation between $(4\pi R^3
\rho_{oc}/\xi_I^3)$ and the total rest mass $M_o$.  From our knowledge
of Newtonian polytropes we expect ($4\pi
R^3\rho_{oc}/\xi_I^3)=M_o/(\xi_1^2|\theta_1'|)+{\cal O}(q_c)$, and we
can obtain a relation accurate to PN order by evaluating $M_o$:
\begin{eqnarray}
M_o & = & \int_0^{M_o}dm' \\
& = & \int_0^R\rho_o\left(1-{2Gm \over rc^2}\right)^{-1/2}4\pi r^2\,dr \\
& = & {4\pi R^3\rho_{oc}\over \xi_I^3}
\int_0^{\xi_I}(\theta+q_c\alpha)^n\left(1+{Gm \over
rc^2}\right)\xi^2\,d\xi\\
& = & {4\pi R^3\rho_{oc}\over \xi_I^3}
\int_0^{\xi_I}\theta^n\left(1+n{q_c\alpha\over\theta}+{4\pi
G\rho_{oc}R^2\xi|\theta'|\over \xi_I^2c^2}\right)\xi^2\,d\xi \\
& = & {4\pi R^3\rho_{oc}\over \xi_I^3}
\left(\int_0^{\xi_I}\theta^n\xi^2\,d\xi
+q_c\int_o^{\xi_I}\left(n\alpha\theta^{n-1}\xi^2-(n+1)\theta^n\theta'\xi^3\right)\,d\xi\right).
\label{aofm}
\end{eqnarray}
Making use of
\begin{equation}
\int_0^{\xi_I}\theta^n\xi^2\,d\xi=\xi_1^2|\theta_1'|+{\cal
O}(q_c^{n+1}),~~~
\int_0^{\xi_I}\theta^n\theta'\xi^3\,d\xi={-3 \over
5-n}\xi_1^3|\theta_1'|^2+{\cal O}(q_c^{n+1}), \label{le.relations}
\end{equation}
in equation (\ref{aofm}) gives
\begin{eqnarray}
{4\pi R^3\rho_{oc}\over \xi_I^3}
& = & {M_o \over \xi_1^2|\theta_1'|}\left(1-{nq_c\over
\xi_1^2|\theta_1'|}\int_0^{\xi_I}\alpha\theta^{n-1}\xi^2\,d\xi
-{3(n+1)\over 5-n}q_c\xi_1|\theta_1'|\right)
\label{almost.final.aofm}\\
& = & {M_o \over \xi_1^2|\theta_1'|}\left(1+q_c{\beta_1\over
|\theta_1'|} +q_c{(n-3)(n+1)\over 5-n}\xi_1|\theta_1'|\right),
\label{final.aofm}
\end{eqnarray}
where $\beta_1=\beta(\xi_1)$.  We evaluated the integral in equation
(\ref{almost.final.aofm}) with the help of equations (\ref{dbeta}) and
(\ref{justified}).  Equation (\ref{final.aofm}) is our desired relation
between $(4\pi R^3\rho_{oc}/\xi_I^3)$ and $M_o$.

Using equations (\ref{final.aofm}) and (\ref{final.udm}) in equation
(\ref{udmprime}) gives
\begin{eqnarray}
\int_0^{M_o}u\,dm'
& = & k_1K\rho_{oc}^{1/n}M_o\left(1+{n(n+1)q_c\over
k_1\xi_1^2|\theta_1'|}\int_0^{\xi_1}\alpha\theta^n\xi^2\,d\xi
+q_c{\beta_1\over |\theta_1'|} \right. \nonumber\\
& & \left. +q_c{(n-3)(n+1)\over 5-n}\xi_1|\theta_1'|\right)-I_1.
\label{final.udmprime}
\end{eqnarray}
We note that the lowest order term is the usual internal energy
contribution for a pure polytrope, $k_1K\rho_{oc}^{1/n}M_o$, as
expected; the first order corrections will help give us our expression
for the coefficient $l_1$.  Using equation (\ref{final.udmprime}) and
the definition of $k_1\equiv n(n+1)\xi_1|\theta_1'|/(5-n)$ in equation
(\ref{upn.app}) gives
\begin{eqnarray}
U_{PN} & = &
{K\rho_{oc}^{1/n}M_on(n+1)q_c\over\xi_1^2|\theta_1'|}\left(\int_0^{\xi_1}\alpha\theta^n\xi^2\,d\xi
+{\beta_1 \xi_1^3 |\theta_1'|\over (5-n)} \right. \nonumber \\
&&\left.+{(n-3)(n+1)\over(5-n)^2}\xi_1^4|\theta_1'|^3\right)+I_3,
\end{eqnarray}
or, after using equations (\ref{qc.ov-based}) and (\ref{final.aofm}) to
lowest order and simplifying,
\begin{eqnarray}
l_1 & = & {(4\pi)^{1/3}\over(\xi_1^2|\theta_1'|)^{5/3}}n
\left(
-\int_0^{\xi_1}\alpha\theta^n\xi^2\,d\xi
-{\beta_1 \xi_1^3 |\theta_1'|\over (5-n)} \right. \nonumber \\
&&\left.
-{(n-3)(n+1)\over(5-n)^2}\xi_1^4|\theta_1'|^3
+\int_0^{\xi_1}d\xi\,\theta^n\xi\int_0^\xi\theta^{n+1}\xi^2\,d\xi
\right).
\label{almost.final.l1}
\end{eqnarray}
The relation\footnotemark \footnotetext{To prove this relation, begin
by non-dimensionalizing the double integral $I_3$ and writing it in
terms of the single integrals $I_1$ and $I_2$ in non-dimensional form,
following the strategy of Exercise 6.19 in Ref. \cite{st83}.  Then with
repeated integration by parts and use of the Lane-Emden equation, the
two remaining integrals can be written in terms of each other, and the
identity is obtained.}
\begin{equation}
\int_0^{\xi_1}d\xi\,\theta^n\xi\int_0^\xi\theta^{n+1}\xi^2\,d\xi =
{n+1\over 5-n}\xi_1^4|\theta_1'|^3
-{n-1\over 3}\int_0^{\xi_1}\xi^3\theta'\theta^{n+1}\,d\xi
\end{equation}
can be used to reduce the double integral in equation
(\ref{almost.final.l1}) to a single integral:
\begin{eqnarray}
l_1 & = & {(4\pi)^{1/3}\over(\xi_1^2|\theta_1'|)^{5/3}}n\left(
-{n-1\over 3}\int_0^{\xi_1}\xi^3\theta'\theta^{n+1}\,d\xi
\right. \nonumber \\
&&\left.-\int_0^{\xi_1}\alpha\theta^n\xi^2\,d\xi
-{\beta_1 \xi_1^3 |\theta_1'|\over 5-n}
-2{(n-4)(n+1)\over (5-n)^2}\xi_1^4|\theta_1'|^3\right). \label{final.l1}
\end{eqnarray}
If desired, the identity [which can be proved from equations
(\ref{dalpha}) and (\ref{dbeta}) and appropriate integrations by parts]
\begin{equation}
\int_0^{\xi_1}\alpha\theta^n\xi^2\,d\xi={1\over 1-n}
\left[
\alpha_1\xi_1^2|\theta_1'|
-{1\over 6}
\left(n^2+25n+28\right)\int_0^{\xi_1}\xi^3\theta'\theta^{n+1}\,d\xi
\right]
\end{equation}
could be used to further manipulate equation (\ref{final.l1}) when
$n\neq 1$, while for $n=1$ this integral can be evaluated using
equation (\ref{alpha.n=1}).  However, we find it more convenient to
work with equation (\ref{final.l1}) directly.

Equation (\ref{final.l1}) is our final expression for the coefficient
$l_1$, and was obtained by evaluating equation (\ref{upn.app}) with the
OV-based trial density functions defined in equation (\ref{ov-based}).
We could proceed to calculate $l_2$ from equation (\ref{wpn.app}) in a
similar manner.  However, it is more straightforward to take the
following approach.

Rearranging equations (\ref{define.epn}) and (\ref{deltaeself}) yields
the PN accurate energy functional for an isolated spherical star, not
necessarily in equilibrium, obeying a polytropic equation of state:
\begin{equation}
E=k_1K\rho_{oc}^{1/n}M_o-k_2G\rho_{oc}^{1/3}M_o^{5/3} -l_1{G \over
c^2}K \rho_{oc}^{{1\over n}+{1\over 3}}M_o^{5/3}
	   -l_2{G^2 \over c^2}\rho_{oc}^{2/3}M_o^{7/3}.
	   \label{isolated.energy.functional}
\end{equation}
Along an {\it equilibrium} sequence we have $0=dE/d\rho_{oc}$, which
gives, to lowest order,
\begin{equation}
K\rho_{oc}^{1/n}={nk_2\over 3 k_1}GM_o^{2/3}\rho_{oc}^{1/3}.
\label{hydro.equil}
\end{equation}
Indeed, equation (\ref{hydro.equil}) is the same relation obtained by
writing the Newtonian equation of hydrostatic equilibrium in terms of
Lane-Emden functions and simplifying.  To PN order, the energy along an
equilibrium sequence therefore satisfies
\begin{equation}
E^{eq}=k_1K\rho_{oc}^{1/n}M_o-k_2G\rho_{oc}^{1/3}M_o^{5/3}-(l_1{nk_2\over
3 k_1}+l_2){G^2 \over c^2}\rho_{oc}^{2/3}M_o^{7/3}. \label{eeq}
\end{equation}

Since here we are focusing on the equilibrium sequence, the first
variational derivative of $E$ with respect to the density functional
must vanish; therefore, any variation in the density profile of PN
order causes a variation in the energy of order P$^2$N, so that it is
sufficient to use Lane-Emden profiles when evaluating the Newtonian
order integrals in $\Delta E$.  We therefore find that along an
equilibrium sequence $\Delta E=I_1+I_2+I_3+I_4+I_5$.  The Newtonian
equation of hydrostatic equilibrium, equation (\ref{hydro.equil}), can
then be used to evaluate $I_1$ through $I_5$ in this case.
These integrals have been carried out by Ref. \cite{st83}
(see their \S 6.9), so that
\begin{equation}
E^{eq}=k_1K\rho_{oc}^{1/n}M_o-k_2G\rho_{oc}^{1/3}M_o^{5/3}-k{G^2 \over
c^2}\rho_{oc}^{2/3}M_o^{7/3}, \label{st.eeq}
\end{equation}
where (see eq. [6.9.31] of \cite{st83})
\begin{eqnarray}
k & = & {(4 \pi)^{2/3} \over (5-n)[\xi_1^2|\theta_1'|]^{7/3}}\left(
-{5+2n-n^2 \over n+1}2\int_0^{\xi_1}\xi^3\theta'\theta^{n+1}\,d\xi
+{3 \over 2}(n-1)\int_0^{\xi_1}\xi^4\theta'^2\theta^n\,d\xi
\right) \\
& = & {(4 \pi)^{2/3} \over [\xi_1^2|\theta_1'|]^{7/3}}\left(
{n-1\over 2(5-n)}\xi_1^4|\theta_1'|^3
-{(n+2)(n+5) \over 6(n+1)}\int_0^{\xi_1}\xi^3\theta'\theta^{n+1}\,d\xi
\right). \label{k4}
\end{eqnarray}
By integrating the OV equation exactly and extracting coefficients from
the resulting equilibrium sequence energy $E^{eq}(\rho_{oc})$, we have
numerically verified that equation (\ref{st.eeq}) holds to PN order.
We emphasize that equation (\ref{eeq}) or (\ref{st.eeq}), in contrast
to equation (\ref{isolated.energy.functional}), should not be used as
energy functionals in an energy variational treatment, since they are
valid only for equilibrium configurations\footnotemark\footnotetext{
In the special case $n=0$, the self energy correction $U_{PN}=0$, and
the use of equation (\ref{st.eeq}) as an energy functional does yield
correct answers to first PN order.  Equation (\ref{st.eeq}) also yields
PN accurate answers when $n=3$; this is because $U_{PN}$
and $W_{PN}$ both then scale as $\rho_{oc}^{2/3}$ and because the
equilibrium relation between $K$ and $M_o$ is independent of
$\rho_{oc}$ at lowest order [see eq. (\ref{hydro.equil})].  However,
for $n\neq 0$ and $n\neq 3$, use of equation (\ref{st.eeq}) as an
energy functional gives errors at first PN order.};
this has not been noticed in previous analyses.
Comparing equations (\ref{eeq}) and (\ref{st.eeq}) we see immediately
\begin{eqnarray}
l_2 & = & k-l_1{nk_2\over 3k_1} \\
    & = & k-{l_1\over
    n+1}\left({4\pi\over\xi_1^4|\theta_1'|^2}\right)^{1/3} \\
& = & 
{(4 \pi)^{2/3} \over [\xi_1^2|\theta_1'|]^{7/3}}\left(
{n-10\over 6}\int_0^{\xi_1}\xi^3\theta'\theta^{n+1}\,d\xi
\right. \nonumber \\
&&\left.
+{n\over n+1}\int_0^{\xi_1}\alpha\theta^n\xi^2\,d\xi
+{n \beta_1 \xi_1^3 |\theta_1'|\over (n+1)(5-n)}
+{3n^2-10n-5\over2(5-n)^2}\xi_1^4|\theta_1'|^3\right). \label{final.l2}
\end{eqnarray}

Equations (\ref{final.l1}) and (\ref{final.l2}) are our final
expressions for the coefficients $l_1$ and $l_2$, and were used to
generate the long dashed curves in Figure~\ref{isolated.spherical}.
The Newtonian curves in this figure, as well as the $n=1$ curves which
use equation (\ref{st.eeq}), have a maximum in the gravitational mass
$M$ which is not accompanied by a maximum in the rest mass $M_o$; this
is because along these sequences $dM/d\rho_{oc}=(\partial M/\partial
M_o)(dM_o/d\rho_{oc})$ vanishes due to $\partial M/\partial M_o=0$.
The remaining sequences have maxima in $M$ and $M_o$ at the same
$\rho_{oc}$, where $0=dM/d\rho_{oc}=dM_o/d\rho_{oc}$, and these turning
points mark the onset of radial instability.

The derivation of the coefficients $l_1$ and $l_2$ is formally invalid
if $n=0$, since then the first relation in equation
(\ref{le.relations}) is accurate only to Newtonian order.  However, for
$n=0$ one can analytically derive the coefficients exactly to be
$l_1=0$ and $l_2=3(4\pi/3)^{2/3}/70$, and the numerically obtained
values given by equations (\ref{final.l1}) and (\ref{final.l2}) do
indeed approach these limiting values as $n$ approaches zero.  See
Table \ref{structureconstants} for numerical values of these
coefficients for various $n$.  We note that the coefficient $l_1$ is
fairly well approximated simply by $l_1\approx n$.  Also note that for
$n\gtrsim 0.5$, $l_1$ is roughly 10 or more times larger in magnitude
than $l_2$, so that we expect $U_{PN}$ to be significantly larger than
$W_{PN}$ in magnitude for all but the stiffest equations of state.

%
%
%
%

%
%

\begin{table}
\caption{\label{structureconstants}The polytropic coefficients to the
PN energy corrections of an isolated spherical star, as determined from
equations (\ref{final.l1}) and (\ref{final.l2}).}
\begin{tabular}{ccr}
$n$& $l_1$ & $l_2$\\
\tableline
0.0&0.000000& 0.111365\\
0.1&0.094949& 0.097502\\
0.2&0.191634& 0.084840\\
0.3&0.289737& 0.073070\\
0.4&0.388996& 0.061974\\
0.5&0.489198& 0.051391\\
0.6&0.590166& 0.041201\\
0.7&0.691752& 0.031312\\
0.8&0.793835& 0.021649\\
0.9&0.896311& 0.012154\\
1.0&0.999096& 0.002779\\
1.1&1.102118&-0.006516\\
1.2&1.205321&-0.015767\\
1.3&1.308657&-0.025003\\
1.4&1.412088&-0.034252\\
1.5&1.515586&-0.043538\\
2.0&2.033637&-0.091306\\
2.5&2.553451&-0.143291\\
3.0&3.080363&-0.202770\\
3.5&3.628218&-0.275041\\
4.0&4.231742&-0.371100\\
4.5&5.003087&-0.523069\\
\end{tabular}
\end{table}

\begin{table}
\caption{\label{isolated.spherical.models} Isolated spherical
polytropic models, as calculated from the PN-accurate energy functional
(\ref{isolated.energy.functional}).  Here $n$ is the polytropic index;
$K\rho_{oc}^{1/n}/c^2$ is the central density parameter; $M$ is the
gravitational mass; $R$ is the stellar radius in Schwarzschild
coordinates as calculated from eq.  (\ref{rhoofr}); $M_o$ is the rest
mass; and $R_o$ is the Newtonian radius defined by eq. (\ref{ro}).  The
value of $K\rho_{oc}/c^2$ marked by an asterisk (*) identifies the star
with maximum rest mass $M_o$ and gravitational mass $M$ for that value
of $n$.}
\begin{tabular}{cddccd}
$n$& log $K\rho_{oc}/c^2$ & $GM/Rc^2$ & $GM_o/R_oc^2$ & $M/M_o$ & $M_o/(c^{3-n}G^{-3/2}K^{n/2})$ \\
\tableline
%
%
%
0.5&     -1.0693&  0.125 &      0.1234&    0.9276&    0.0580\\
0.5&     -1.0076&  0.1380&      0.1359&    0.9198&    0.0655\\ 
0.5&     -0.7687&  0.1941&      0.1883&    0.8853&    0.0984\\ 
0.5&     -0.7450&  0.2   &      0.1936&    0.8816&    0.1019\\
0.5&     -0.5541&  0.2481&      0.2348&    0.8517&    0.1297\\ 
0.5&     -0.5464&  0.25  &      0.2364&    0.8505&    0.1308\\
0.5&      0.1212*&  0.3917&      0.3058&    0.7875&    0.1804\\
\\
1.0&     -1.9749&  0.02  &      0.0196&    0.9900&    0.0246\\
1.0&     -1.0180&  0.125 &      0.1074&    0.9387&    0.1346\\
1.0&     -0.8907&  0.15  &      0.1232&    0.9275&    0.1545\\
1.0&     -0.6532&  0.2   &      0.1467&    0.9083&    0.1838\\
1.0&     -0.4171&  0.25  &      0.1563&    0.8984&    0.1960\\
1.0&     -0.3902*&  0.2556&      0.1564&    0.8983&    0.1961\\
\end{tabular}
\end{table}

\begin{table}
\squeezetable
\caption{Irrotational Equilibrium Sequences.  The dimensionless
quantities $\bar \Omega$, $\bar J$, and $\bar E$ are defined in
equation (\ref{dimensions}).  \label{irrotationalwill}}
\begin{tabular}{ccccrccccc}
$r/a_1$ & $r/R_o$ & $rc^2/Gm$ &
$a_2/a_1$ & $a_3/a_1$ & $T_s/|W|$ &
$\bar\Omega$ & $\bar J$ & $\bar E$ &
$R/R_o$\\
\multicolumn{10}{c}{$GM/Rc^2=0$ (Newtonian treatment), $n=0.5$}\\
\tableline
   6.0& 6.052&$\infty$&0.9871&0.9874&   0.371(-6)&0.1097& 1.7401&-1.1937&1.0000\\
   5.0& 5.075&$\infty$&0.9776&0.9783&   0.193(-5)&0.1430& 1.5944&-1.2095&1.0000\\
   4.0& 4.120&$\infty$&0.9557&0.9579&   0.146(-4)&0.1958& 1.4392&-1.2320&1.0002\\
   3.5& 3.661&$\infty$&0.9331&0.9378&   0.492(-4)&0.2343& 1.3602&-1.2467&1.0004\\
   3.2& 3.396&$\infty$&0.9117&0.9193&   0.111(-3)&0.2629& 1.3145&-1.2565&1.0006\\
   3.0& 3.228&$\infty$&0.8921&0.9027&   0.200(-3)&0.2846& 1.2859&-1.2633&1.0009\\
   2.8& 3.068&$\infty$&0.8663&0.8814&   0.373(-3)&0.3084& 1.2603&-1.2698&1.0014\\
   2.6& 2.920&$\infty$&0.8322&0.8537&   0.724(-3)&0.3340& 1.2401&-1.2754&1.0023\\
2.344*& 2.759&$\infty$&0.7716&0.8049&   0.178(-2)&0.3681& 1.2283&-1.2790&1.0042\\
   2.0& 2.623&$\infty$&0.6498&0.7045&   0.635(-2)&0.4090& 1.2651&-1.2663&1.0107\\
\hline
\multicolumn{10}{c}{$GM/Rc^2=0.2$, $n=0.5$}\\
\hline
   6.0& 5.173& 13.36&0.9845&0.9848&   0.424(-6)&0.1146& 1.5553&-1.2860&0.8533\\
   5.0& 4.344& 11.22&0.9732&0.9737&   0.210(-5)&0.1458& 1.4727&-1.2953&0.8534\\
   4.0& 3.537&  9.14&0.9475&0.9495&   0.146(-4)&0.1930& 1.3956&-1.3064&0.8536\\
   3.5& 3.150&  8.14&0.9221&0.9259&   0.457(-4)&0.2260& 1.3647&-1.3120&0.8539\\
   3.2& 2.928&  7.56&0.8989&0.9049&   0.972(-4)&0.2499& 1.3520&-1.3146&0.8542\\
   3.0& 2.787&  7.20&0.8784&0.8866&   0.166(-3)&0.2679& 1.3477&-1.3155&0.8547\\
2.922*& 2.734&  7.06&0.8690&0.8782&   0.207(-3)&0.2753& 1.3473&-1.3156&0.8549\\
   2.8& 2.652&  6.85&0.8524&0.8636&   0.293(-3)&0.2875& 1.3485&-1.3153&0.8553\\
   2.6& 2.527&  6.53&0.8192&0.8345&   0.534(-3)&0.3090& 1.3566&-1.3132&0.8563\\
   2.0& 2.258&  5.83&0.6525&0.6899&   0.396(-2)&0.3794& 1.4748&-1.2770&0.8653\\
 1.736& 2.245&  5.80&0.5351&0.5835&   0.103(-1)&0.4036& 1.6258&-1.2255&0.8772\\
\hline
\multicolumn{10}{c}{$GM/Rc^2=0.25$, $n=0.5$}\\
\hline
   6.0& 4.878& 10.32&0.9836&0.9838&   0.437(-6)&0.1171& 1.4952&-1.3223&0.8042\\
   5.0& 4.099&  8.67&0.9716&0.9721&   0.213(-5)&0.1479& 1.4296&-1.3297&0.8043\\
   4.0& 3.340&  7.07&0.9446&0.9465&   0.144(-4)&0.1938& 1.3749&-1.3377&0.8045\\
   3.5& 2.978&  6.30&0.9182&0.9218&   0.444(-4)&0.2256& 1.3588&-1.3406&0.8048\\
3.280*& 2.825&  5.98&0.9014&0.9064&   0.760(-4)&0.2421& 1.3566&-1.3410&0.8051\\
   3.2& 2.770&  5.86&0.8943&0.8998&   0.931(-4)&0.2486& 1.3570&-1.3409&0.8053\\
   3.0& 2.638&  5.58&0.8733&0.8808&   0.157(-3)&0.2658& 1.3612&-1.3400&0.8057\\
   2.8& 2.512&  5.31&0.8470&0.8571&   0.274(-3)&0.2847& 1.3717&-1.3375&0.8065\\
   2.6& 2.396&  5.07&0.8136&0.8274&   0.493(-3)&0.3053& 1.3913&-1.3324&0.8076\\
   2.0& 2.145&  4.54&0.6490&0.6822&   0.352(-2)&0.3748& 1.5575&-1.2825&0.8174\\
 1.682& 2.146&  4.54&0.5069&0.5517&   0.112(-1)&0.4034& 1.7850&-1.2052&0.8341\\
\hline
\\
\\
\\
\\
\hline
\multicolumn{10}{c}{$GM/Rc^2=0.2$, $n=1.0$}\\
\hline
   6.0& 4.026& 13.72&0.9894&0.9895&   0.144(-6)&0.1713& 1.4318&-1.3369&0.6663\\
   5.0& 3.373& 11.50&0.9816&0.9819&   0.719(-6)&0.2188& 1.3528&-1.3502&0.6663\\
   4.0& 2.732&  9.31&0.9638&0.9648&   0.506(-5)&0.2917& 1.2766&-1.3668&0.6667\\
   3.5& 2.422&  8.25&0.9459&0.9480&   0.161(-4)&0.3434& 1.2431&-1.3759&0.6672\\
   3.2& 2.241&  7.64&0.9294&0.9327&   0.346(-4)&0.3813& 1.2266&-1.3811&0.6679\\
   3.0& 2.125&  7.24&0.9147&0.9193&   0.598(-4)&0.4098& 1.2182&-1.3840&0.6686\\
   2.8& 2.013&  6.86&0.8957&0.9022&   0.107(-3)&0.4413& 1.2127&-1.3860&0.6698\\
2.660*& 1.939&  6.61&0.8792&0.8874&   0.163(-3)&0.4651& 1.2115&-1.3865&0.6710\\
   2.6& 1.908&  6.50&0.8712&0.8803&   0.197(-3)&0.4757& 1.2117&-1.3864&0.6717\\
   2.0& 1.664&  5.67&0.7405&0.7667&   0.159(-2)&0.5884& 1.2636&-1.3615&0.6889\\
 1.721& 1.636&  5.58&0.6317&0.6720&   0.481(-2)&0.6266& 1.3521&-1.3146&0.7143\\
\hline
\multicolumn{10}{c}{$GM/Rc^2=0.25$, $n=1.0$}\\
\hline
   6.0& 3.398& 10.87&0.9890&0.9891&   0.149(-6)&0.2128& 1.3490&-1.3952&0.5621\\
   5.0& 2.850&  9.11&0.9808&0.9811&   0.739(-6)&0.2700& 1.2860&-1.4083&0.5626\\
   4.0& 2.316&  7.41&0.9623&0.9633&   0.512(-5)&0.3557& 1.2303&-1.4232&0.5647\\
   3.5& 2.063&  6.60&0.9438&0.9458&   0.160(-4)&0.4144& 1.2099&-1.4299&0.5675\\
   3.2& 1.919&  6.14&0.9269&0.9301&   0.342(-4)&0.4559& 1.2024&-1.4327&0.5706\\
   3.0& 1.827&  5.84&0.9118&0.9162&   0.587(-4)&0.4863& 1.2006&-1.4334&0.5737\\
2.987*& 1.822&  5.83&0.9107&0.9152&   0.608(-4)&0.4884& 1.2006&-1.4335&0.5739\\
   2.8& 1.741&  5.57&0.8926&0.8987&   0.104(-3)&0.5189& 1.2023&-1.4327&0.5779\\
   2.6& 1.662&  5.32&0.8678&0.8765&   0.190(-3)&0.5531& 1.2089&-1.4296&0.5836\\
   2.0& 1.500&  4.80&0.7382&0.7628&   0.149(-2)&0.6538& 1.2835&-1.3897&0.6194\\
 1.688& 1.514&  4.84&0.6158&0.6556&   0.515(-2)&0.6787& 1.3948&-1.3250&0.6631\\
\end{tabular}
\end{table}

\newpage

\begin{table}
\squeezetable
\caption{Irrotational Equilibrium Sequences: Energy Terms \label{irrotationale}}
\begin{tabular}{ccccrcccccc}
$r/a_1$ & $\bar U+\bar U'$ & $\bar W+\bar W'$ &  $\bar T_s+\bar T_s'$ &
$\bar W_t$ & $\bar E_H$ & $\bar E_{PN}$ & $\bar U_{PN}+\bar U_{PN}'$ &
$\bar W_{PN}+\bar W_{PN}'$ & $\bar E_{OI}+\bar E_{OI}'$ & $\bar E$\\
\multicolumn{11}{c}{$GM/Rc^2=0.2$, $n=1.0$}\\
\hline
   6.0&  1.5147& -2.1704&  0.0000&  0.0000& -0.1234& -0.0010& -0.5927& -0.0015&  0.0374& -1.3369\\
   5.0&  1.5142& -2.1701&  0.0000& -0.0001& -0.1440& -0.0012& -0.5924& -0.0015&  0.0450& -1.3502\\
   4.0&  1.5121& -2.1689&  0.0000& -0.0005& -0.1714& -0.0015& -0.5913& -0.0015&  0.0562& -1.3668\\
   3.5&  1.5086& -2.1669&  0.0000& -0.0010& -0.1879& -0.0016& -0.5895& -0.0014&  0.0638& -1.3759\\
   3.2&  1.5042& -2.1644&  0.0001& -0.0016& -0.1981& -0.0017& -0.5872& -0.0014&  0.0692& -1.3811\\
   3.0&  1.4993& -2.1616&  0.0001& -0.0023& -0.2047& -0.0018& -0.5847& -0.0014&  0.0731& -1.3840\\
   2.8&  1.4917& -2.1572&  0.0002& -0.0034& -0.2106& -0.0018& -0.5807& -0.0014&  0.0774& -1.3860\\
2.660*&  1.4839& -2.1526&  0.0004& -0.0045& -0.2140& -0.0019& -0.5767& -0.0014&  0.0804& -1.3865\\
   2.6&  1.4797& -2.1502&  0.0004& -0.0051& -0.2151& -0.0019& -0.5745& -0.0014&  0.0817& -1.3864\\
   2.0&  1.3752& -2.0864&  0.0033& -0.0182& -0.2027& -0.0027& -0.5211& -0.0014&  0.0925& -1.3615\\
 1.696&  1.2219& -1.9844&  0.0106& -0.0349& -0.1615& -0.0041& -0.4451& -0.0013&  0.0906& -1.3082\\
\end{tabular}
\end{table}

\begin{table}
\squeezetable
\caption{Irrotational Equilibrium Sequences: Angular Momentum Terms \label{irrotationalj}}
\begin{tabular}{cccccc}
$r/a_1$ & $\bar J_H$ & $\bar J_s+\bar
J_s'$ & $\bar J_{PN}$ & $\bar J_{OI}+\bar
J_{OI}'$ &
$\bar J$ \\
\multicolumn{6}{c}{$GM/Rc^2=0.2$, $n=1.0$}\\
\hline
   6.0&  1.5437&  0.0000& -0.0017& -0.1102&  1.4318\\
   5.0&  1.4524&  0.0000& -0.0008& -0.0988&  1.3528\\
   4.0&  1.3623&  0.0001&  0.0006& -0.0864&  1.2766\\
   3.5&  1.3213&  0.0002&  0.0015& -0.0799&  1.2431\\
   3.2&  1.3000&  0.0005&  0.0022& -0.0760&  1.2266\\
   3.0&  1.2882&  0.0007&  0.0026& -0.0734&  1.2182\\
   2.8&  1.2794&  0.0012&  0.0030& -0.0709&  1.2127\\
2.660*&  1.2757&  0.0017&  0.0032& -0.0692&  1.2115\\
   2.6&  1.2749&  0.0021&  0.0033& -0.0685&  1.2117\\
   2.0&  1.3112&  0.0130&  0.0032& -0.0638&  1.2636\\
 1.696&  1.3890&  0.0388&  0.0008& -0.0647&  1.3638\\
\end{tabular}
\end{table}

\begin{table}
\squeezetable
\caption{Equilibrium Corotational Sequences.  The dimensionless
quantities $\bar \Omega$, $\bar J$, $\bar E$, and $\bar {\cal C}$ are
defined in equation (\ref{dimensions}).  \label{corotatingwill}}
\begin{tabular}{cccccccccccc}
$r/a_1$ & $r/R_o$ & $rc^2/Gm$ & $a_2/a_1$ & $a_3/a_1$ & $T_s/|W|$ & $\bar\Omega$ & $\bar J$ & $\bar E$ & $R/R_o$ & $\bar {\cal C}$\\
\multicolumn{11}{c}{$GM/Rc^2=0$ (Newtonian treatment), $n=0.5$}\\
\tableline
   6.0& 6.073&$\infty$&0.9874&0.9794&   0.221(-2)&0.1092& 1.8057&-1.1904&1.0009&-0.0623\\
   5.0& 5.105&$\infty$&0.9784&0.9651&   0.376(-2)&0.1418& 1.6809&-1.2039&1.0016&-0.0816\\
   4.0& 4.162&$\infty$&0.9585&0.9348&   0.711(-2)&0.1929& 1.5603&-1.2211&1.0031&-0.1130\\
   3.5& 3.710&$\infty$&0.9390&0.9065&   0.103(-1)&0.2299& 1.5072&-1.2307&1.0046&-0.1369\\
   3.2& 3.450&$\infty$&0.9214&0.8819&   0.131(-1)&0.2571& 1.4810&-1.2363&1.0061&-0.1554\\
   3.0& 3.283&$\infty$&0.9058&0.8609&   0.156(-1)&0.2777& 1.4674&-1.2394&1.0074&-0.1701\\
   2.8& 3.124&$\infty$&0.8860&0.8352&   0.186(-1)&0.3005& 1.4583&-1.2417&1.0092&-0.1871\\
2.625*& 2.992&$\infty$&0.8641&0.8079&   0.219(-1)&0.3221& 1.4553&-1.2425&1.0112&-0.2042\\
   2.6& 2.974&$\infty$&0.8605&0.8035&   0.224(-1)&0.3254& 1.4554&-1.2424&1.0115&-0.2068\\
2.115**&2.677&$\infty$&0.7609&0.6909&   0.367(-1)&0.3915& 1.4949&-1.2298&1.0218&-0.2696\\
   2.0& 2.630&$\infty$&0.7256&0.6543&   0.416(-1)&0.4066& 1.5208&-1.2208&1.0259&-0.2880\\
\hline
\multicolumn{11}{c}{$GM/Rc^2=0.2$, $n=0.5$}\\
\hline
 6.0   & 5.189& 13.40&0.9848&0.9782&   0.175(-2)&0.1141& 1.6096&-1.2832&0.8541&-0.0524\\
 5.0   & 4.365& 11.27&0.9739&0.9634&   0.284(-2)&0.1449& 1.5423&-1.2907&0.8546&-0.0671\\
 4.0   & 3.565&  9.21&0.9499&0.9324&   0.503(-2)&0.1911& 1.4893&-1.2983&0.8558&-0.0899\\
 3.5   & 3.182&  8.22&0.9269&0.9039&   0.699(-2)&0.2232& 1.4756&-1.3007&0.8571&-0.1067\\
 3.345*& 3.067&  7.92&0.9172&0.8920&   0.780(-2)&0.2348& 1.4746&-1.3009&0.8576&-0.1129\\
 3.2   & 2.962&  7.65&0.9065&0.8792&   0.867(-2)&0.2465& 1.4756&-1.3007&0.8582&-0.1194\\
 3.0   & 2.821&  7.29&0.8887&0.8584&   0.101(-1)&0.2640& 1.4809&-1.2995&0.8593&-0.1294\\
2.842**& 2.714&  7.01&0.8716&0.8387&   0.115(-1)&0.2790& 1.4893&-1.2975&0.8604&-0.1383\\
 2.8   & 2.687&  6.94&0.8666&0.8329&   0.119(-1)&0.2832& 1.4922&-1.2968&0.8608&-0.1408\\
 2.6   & 2.560&  6.61&0.8388&0.8017&   0.141(-1)&0.3042& 1.5117&-1.2918&0.8627&-0.1542\\
 1.734 & 2.231&  5.76&0.5976&0.5539&   0.357(-1)&0.4054& 1.8183&-1.1935&0.8899&-0.2491\\
\hline
\multicolumn{11}{c}{$GM/Rc^2=0.25$, $n=0.5$}\\
\hline
 6.0   & 4.892& 10.35&0.9839&0.9778&   0.160(-2)&0.1167& 1.5459&-1.3196&0.8050&-0.0492\\
 5.0   & 4.117&  8.71&0.9723&0.9627&   0.257(-2)&0.1470& 1.4941&-1.3255&0.8055&-0.0625\\
 4.0   & 3.365&  7.12&0.9470&0.9312&   0.447(-2)&0.1921& 1.4606&-1.3303&0.8066&-0.0830\\
 3.706*& 3.151&  6.67&0.9341&0.9157&   0.537(-2)&0.2096& 1.4580&-1.3307&0.8073&-0.0912\\
 3.5   & 3.006&  6.36&0.9227&0.9022&   0.615(-2)&0.2232& 1.4596&-1.3304&0.8078&-0.0979\\
3.230**& 2.820&  5.96&0.9038&0.8801&   0.741(-2)&0.2432& 1.4674&-1.3288&0.8089&-0.1079\\
 3.2   & 2.800&  5.92&0.9013&0.8773&   0.757(-2)&0.2456& 1.4688&-1.3285&0.8090&-0.1091\\
 3.0   & 2.668&  5.64&0.8828&0.8562&   0.878(-2)&0.2624& 1.4813&-1.3258&0.8101&-0.1180\\
 2.8   & 2.542&  5.38&0.8598&0.8305&   0.103(-1)&0.2809& 1.5010&-1.3211&0.8116&-0.1282\\
 2.6   & 2.425&  5.13&0.8312&0.7990&   0.122(-1)&0.3012& 1.5304&-1.3137&0.8136&-0.1401\\
 1.679 & 2.131&  4.51&0.5634&0.5264&   0.338(-1)&0.4053& 1.9611&-1.1760&0.8463&-0.2354\\
\hline
\\
\\
\\
\hline
\multicolumn{11}{c}{$GM/Rc^2=0.2$, $n=1.0$}\\
\hline
 6.0   & 4.045& 13.79&0.9895&0.9847&   0.127(-2)&0.1702& 1.4714&-1.3338&0.6683&-0.0373\\
 5.0   & 3.398& 11.58&0.9820&0.9741&   0.208(-2)&0.2166& 1.4040&-1.3450&0.6697&-0.0479\\
 4.0   & 2.769&  9.44&0.9651&0.9516&   0.373(-2)&0.2867& 1.3461&-1.3574&0.6728&-0.0645\\
 3.5   & 2.466&  8.41&0.9486&0.9303&   0.523(-2)&0.3354& 1.3260&-1.3628&0.6759&-0.0769\\
 3.2   & 2.292&  7.81&0.9337&0.9117&   0.652(-2)&0.3706& 1.3195&-1.3647&0.6787&-0.0863\\
 3.056*& 2.211&  7.53&0.9246&0.9005&   0.729(-2)&0.3893& 1.3186&-1.3650&0.6805&-0.0916\\
 3.0   & 2.180&  7.43&0.9206&0.8957&   0.762(-2)&0.3969& 1.3188&-1.3650&0.6813&-0.0938\\
 2.8   & 2.073&  7.06&0.9041&0.8759&   0.899(-2)&0.4255& 1.3220&-1.3638&0.6848&-0.1025\\
 2.6   & 1.972&  6.72&0.8831&0.8512&   0.107(-1)&0.4565& 1.3307&-1.3605&0.6895&-0.1126\\
2.570**& 1.957&  6.67&0.8795&0.8470&   0.110(-1)&0.4613& 1.3326&-1.3597&0.6903&-0.1143\\
 1.730 & 1.703&  5.81&0.6886&0.6420&   0.276(-1)&0.5933& 1.5182&-1.2710&0.7500&-0.1913\\
\hline
\multicolumn{11}{c}{$GM/Rc^2=0.25$, $n=1.0$}\\
\hline
 6.0   & 3.460& 11.06&0.9891&0.9845&   0.122(-2)&0.2077& 1.3903&-1.3908&0.5715&-0.0339\\
 5.0   & 2.926&  9.36&0.9813&0.9738&   0.197(-2)&0.2607& 1.3385&-1.4012&0.5763&-0.0435\\
 4.0   & 2.409&  7.70&0.9638&0.9510&   0.350(-2)&0.3381& 1.2990&-1.4113&0.5849&-0.0585\\
 3.5   & 2.163&  6.92&0.9468&0.9296&   0.487(-2)&0.3901& 1.2895&-1.4142&0.5921&-0.0698\\
 3.371*& 2.101&  6.72&0.9408&0.9222&   0.534(-2)&0.4054& 1.2890&-1.4144&0.5945&-0.0733\\
 3.2   & 2.022&  6.47&0.9315&0.9108&   0.606(-2)&0.4267& 1.2900&-1.4140&0.5982&-0.0784\\
 3.0   & 1.933&  6.18&0.9181&0.8947&   0.707(-2)&0.4534& 1.2942&-1.4124&0.6033&-0.0853\\
2.839**& 1.864&  5.96&0.9049&0.8791&   0.805(-2)&0.4763& 1.3006&-1.4098&0.6083&-0.0917\\
 2.8   & 1.848&  5.91&0.9013&0.8749&   0.832(-2)&0.4820& 1.3026&-1.4090&0.6097&-0.0934\\
 2.6   & 1.769&  5.66&0.8799&0.8502&   0.990(-2)&0.5123& 1.3167&-1.4029&0.6176&-0.1029\\
 1.700 & 1.596&  5.10&0.6739&0.6301&   0.269(-1)&0.6335& 1.5440&-1.2852&0.7053&-0.1832\\
\end{tabular}
\end{table}

\begin{table}
\squeezetable
\caption{\label{compare.to.lw96}Physical quantities at the ISCO for
selected irrotational sequences.  Here $M_{1.4}=M/(1.4 M_\odot)$ and
$M_{1.6}^{\rm max}=M^{\rm max}/(1.6 M_\odot)$, where $M^{\rm max}$ is
the maximum value of the isolated gravitational mass $M$.}
\begin{tabular}{cddccccccccc}
&&&&&&&&&\multicolumn{3}{c}{Ref. \cite{lw96}'s Results}\\ \cline{10-12}
&&&&&&&$M_{1.4}f_{ISCO}$ &$M_{1.6}^{\rm max}f_{ISCO}$ &&&$M_{1.4}f_{ISCO}$ \\
$n$& $GM/Rc^2$& $GM/R_oc^2$ & $Jc/GM_o^2$ & $M_t^2/M_o^2$ & $a_2/a_1$ & $a_3/a_1$ & [Hz] & [Hz] & $a_2/a_1$ & $a_3/a_1$ & [Hz]\\
\tableline
0.5&0.125 & 0.1145&3.7178&3.387& 0.821& 0.839& 258& 596&&&\\
0.5&0.1380& 0.125 &3.5656&3.326& 0.828& 0.845& 289& 597& 0.830& 0.850& 279\\
0.5&0.1941& 0.1667&3.0998&3.072& 0.865& 0.875& 404& 576& 0.857& 0.871& 399\\
0.5&0.2   & 0.1707&3.0622&3.046& 0.869& 0.878& 413& 572&&&\\
0.5&0.2481& 0.2   &2.7997&2.841& 0.900& 0.905& 471& 531& 0.880& 0.891& 488\\
0.5&0.25  & 0.2010&2.7904&2.833& 0.901& 0.906& 473& 529&&&\\
\\
1.0&0.125& 0.1008 &3.6758&3.463& 0.829& 0.846& 297& 362&&&\\
1.0&0.2  & 0.1332 &3.1631&3.228& 0.878& 0.887& 474& 438&&&\\
1.0&0.25 & 0.1405 &3.0364&3.154& 0.911& 0.915& 542& 475&&&\\
\end{tabular}
\end{table}

\begin{table}
\caption{\label{corotating.isco}Orbital frequency at the ISCO for
selected corotating sequences.}
\begin{tabular}{cdcccc}
$n$& $GM/Rc^2$& $M_{1.4}f_{ISCO}^{\rm sec}$ [Hz]& $M_{1.4}f^{\rm dyn}_{ISCO}$ [Hz]& $M_{1.6}^{\rm max}f_{ISCO}^{\rm sec}$ [Hz]& $M_{1.6}^{\rm max}f^{\rm dyn}_{ISCO}$ [Hz] \\
\tableline
0.5&0.125& 220 & 267& 508 &617\\
0.5&0.2  & 353 & 419& 488 &580\\
0.5&0.25 & 409 & 475& 458 &531\\
\\
1.0&0.125& 248 & 300& 302 &366\\
1.0&0.2  & 397 & 471& 367 &434\\
1.0&0.25 & 450 & 529& 394 &463\\
\end{tabular}
\end{table}

%
%

\begin{figure}
\caption{
We plot (a) the rest mass $M_o$ and (b) the gravitational mass
$M=M_o+E/c^2$ for isolated spherical stars as a function of the central
density parameter $K\rho_{oc}^{1/n}/c^2$.  Here mass is in units
$c^{3-n}G^{-3/2}K^{n/2}$.  The solid curve was obtained by integrating
the OV equation without approximation.  The remaining curves were found
by solving the equilibrium equation $0=dE/d\rho_{oc}$ for various
energy functionals $E$.  The dotted-and-dashed curve results from the
use of equation (\ref{st.eeq}) as the energy functional.  The remaining
three curves use different values of the coefficients $l_1$ and $l_2$,
via equation (\ref{isolated.energy.functional}): The dotted curve is
the purely Newtonian result ($l_1=l_2=0$); the short-dashed curve uses
the coefficients from equations (\ref{l1le}) and (\ref{l2le}); the
long-dashed curve uses our standard coefficients from equations
(\ref{final.l1}) and (\ref{final.l2}).  \label{isolated.spherical}
}
\end{figure}

\begin{figure}
\caption{
The total energy $E$, relative to its value $E_\infty$ at infinite
separation, as a function of the orbital frequency $f=\Omega/2\pi$ for
selected $n=0.5$ irrotational sequences.  The thick solid curves
represent our PN sequences with various $GM/Rc^2$, where $M$ and $R$
are the isolated neutron star gravitational mass and radius in
Schwarzschild coordinates, respectively.  The dashed curves represent
purely Newtonian results, with the thin curve corresponding to the
point mass sequence and the thick curve corresponding to the EFE
treatment.  The minima of these curves mark the ISCO, inside of which
the orbit is dynamically unstable.  The sequences terminate when the
stars touch.  \label{evsw2irrotational}
}
\end{figure}

\begin{figure}
\caption{
Same as Fig. \ref{evsw2irrotational}, but for an $n=1$ irrotational
sequence.
\label{evsw2irrotational.n=1}
}\end{figure}

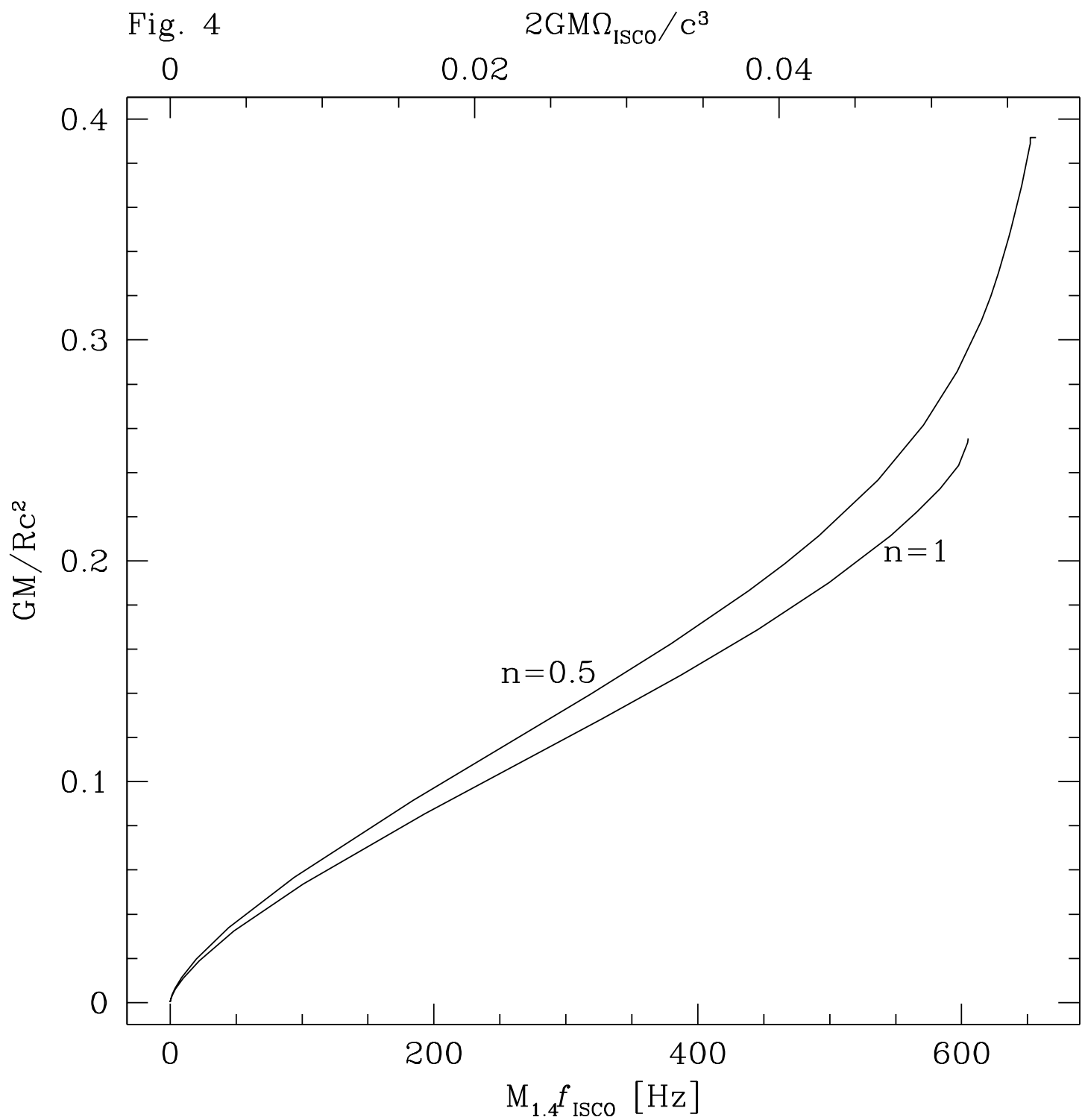
\begin{figure}
\caption{The dependence of $GM/Rc^2$ (as calculated in isolation) on
the orbital frequency $f_{ISCO}$ at the ISCO for $n=0.5$ and $n=1$
irrotational sequences.  For each $n$, the curves are terminated at the
value $GM/Rc^2$ corresponding to the maximum mass.
\label{eosnew}
}
\end{figure}

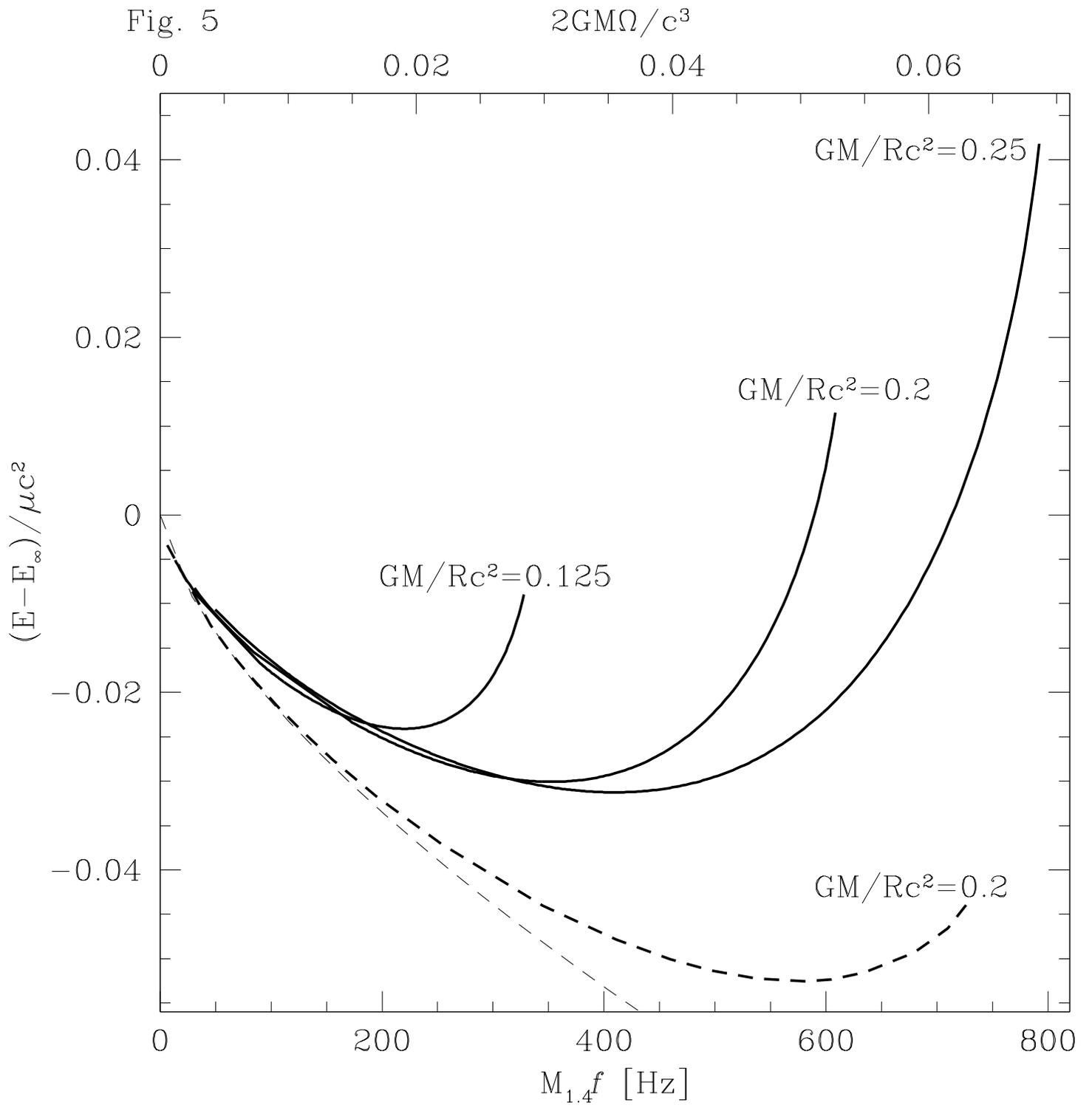
\begin{figure}
\caption{
Same as Fig. \ref{evsw2irrotational}, but for an $n=0.5$ corotating
sequence.  The minima of these curves mark the onset of the secular
instability in the orbit.
\label{evsw2corotating}
}\end{figure}

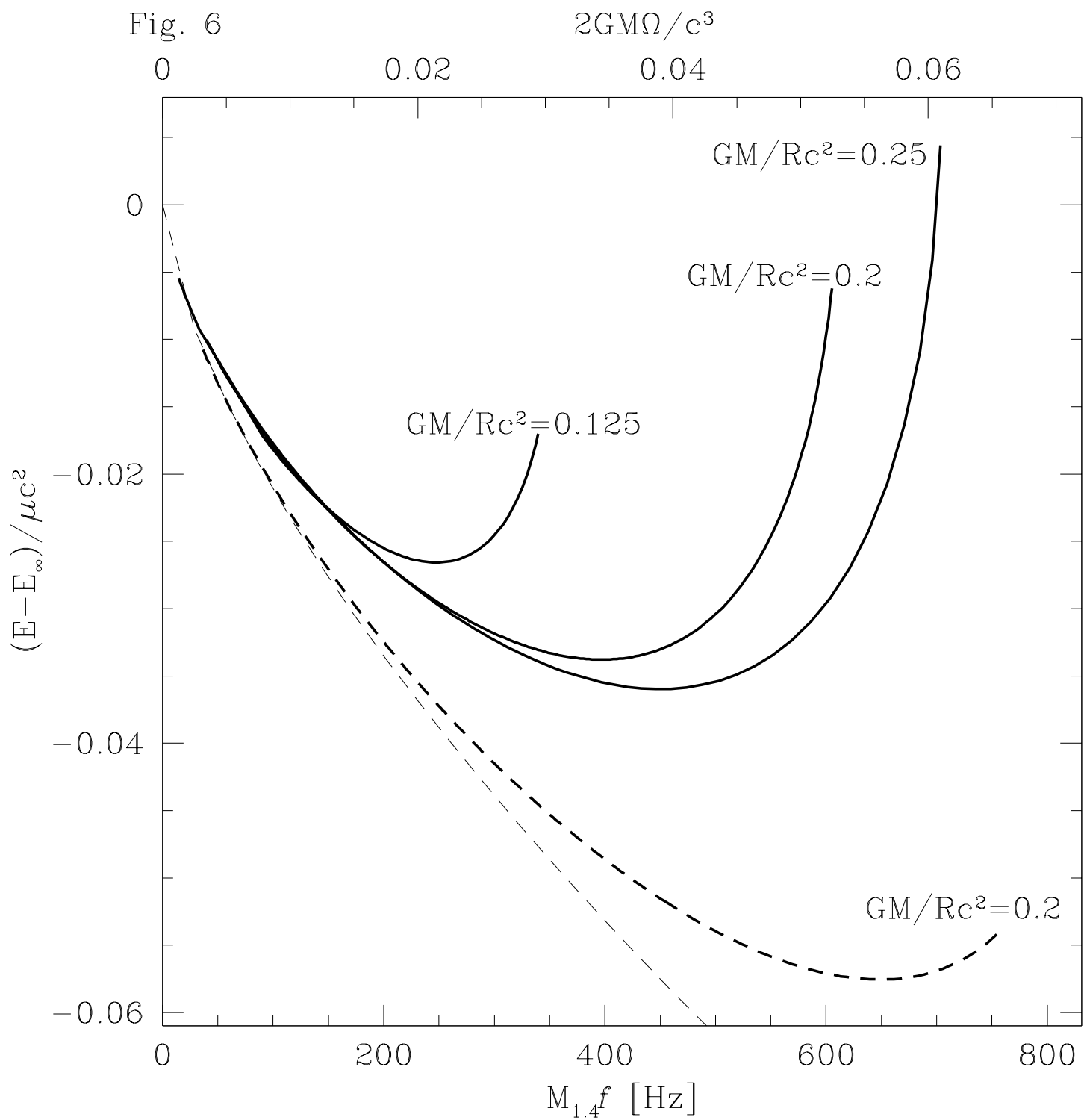
\begin{figure}
\caption{
Same as Fig. \ref{evsw2corotating}, but for an $n=1$ corotating
sequence.
\label{evsw2corotating.n=1}
}\end{figure}

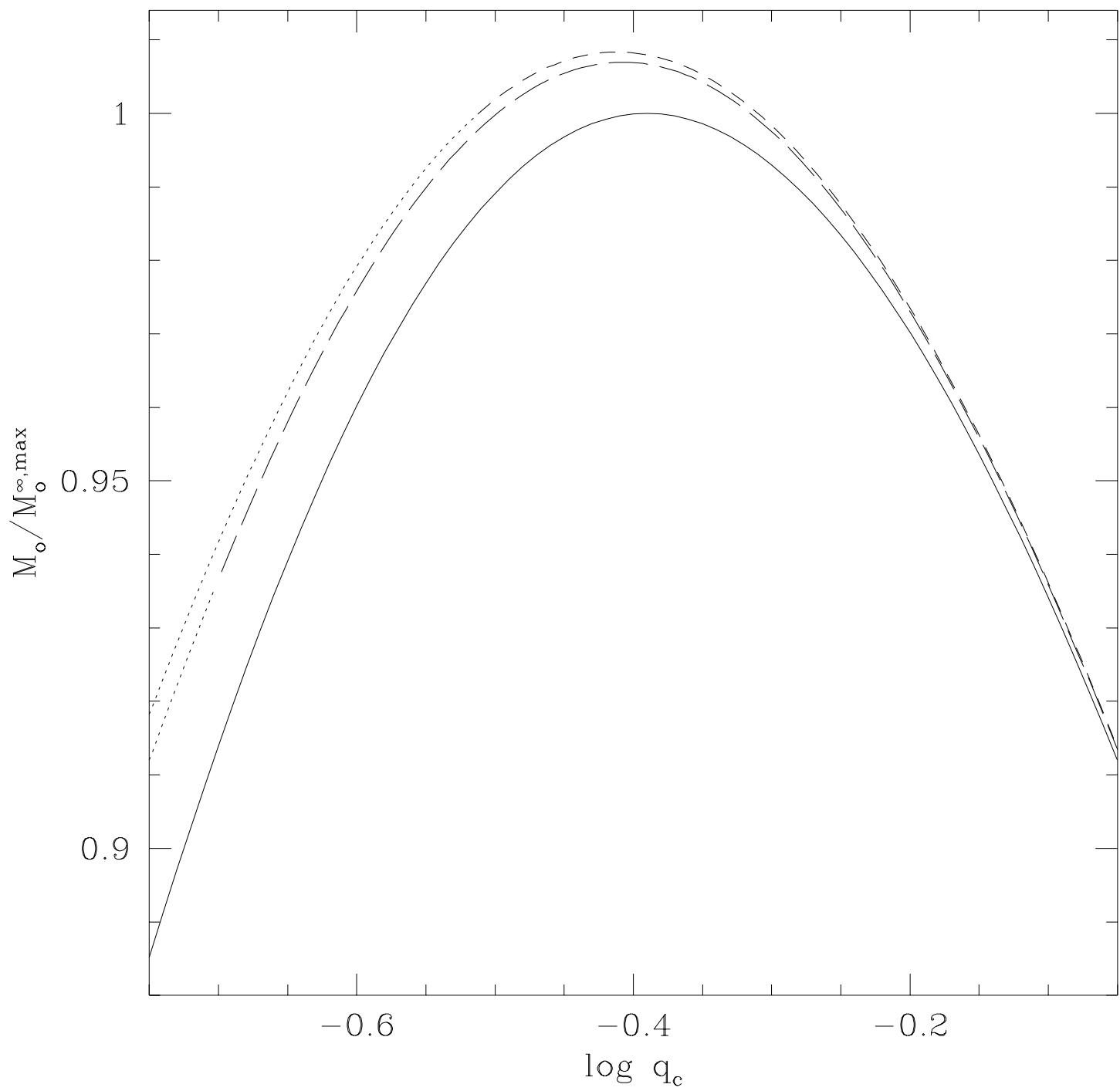
\begin{figure}
\caption{
The rest mass $M_o$, normalized to the maximum rest mass of a star at
infinite separation $M_o^{\infty, \rm max}$, as a function of the
central density parameter $q_c=K \rho_{oc}^{1/n}/c^2$ for an $n=1$
corotating binary system and at three different values of the orbital
angular frequency $f$: $f=0$ (solid curve), $M_{1.6}^{max} f=434$ Hz
(long dashed curve), and $M_{1.6}^{max} f=463$ Hz (short dashed
curve).  The dotted continuations of the dashed curves represent those
binaries inside the innermost dynamically stable orbit.
\label{variousf}
}
\end{figure}

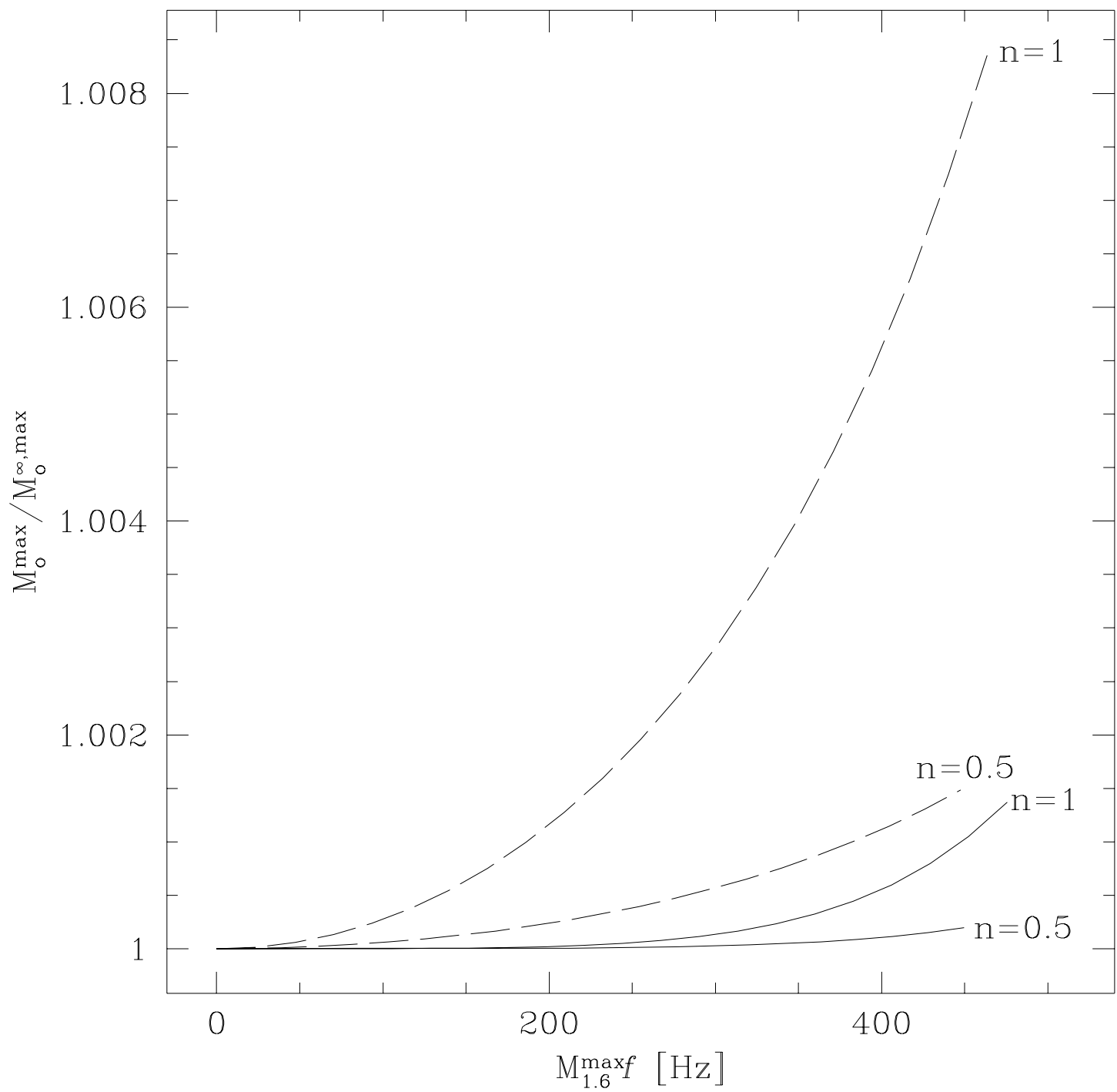
\begin{figure}
\caption{
The maximum equilibrium rest mass $M_o^{\rm max}$ as a function of the
orbital angular frequency $f$ for corotating (dashed curves) and
irrotational (solid curves) sequences.  The curves are terminated at
the values of $f$ for which stars of rest mass $M_o=M_o^{\infty, \rm
max}$ acquire a dynamically unstable orbit.
\label{mvsomega4}
}
\end{figure}

\begin{figure}
\caption{
As in Fig. \ref{evsw2corotating.n=1}, we plot the binding energy
versus orbital frequency $f$ for various $n=1$ corotating sequences:
$\log q_c=-1.15$ (solid curve), $-0.87$ (dashed curve), and $-0.59$
(dotted curve).  The data points represented by crosses are from the
computations of Baumgarte et al., for the
same values of $q_c$.
\label{compare.with.tom}
}
\end{figure}

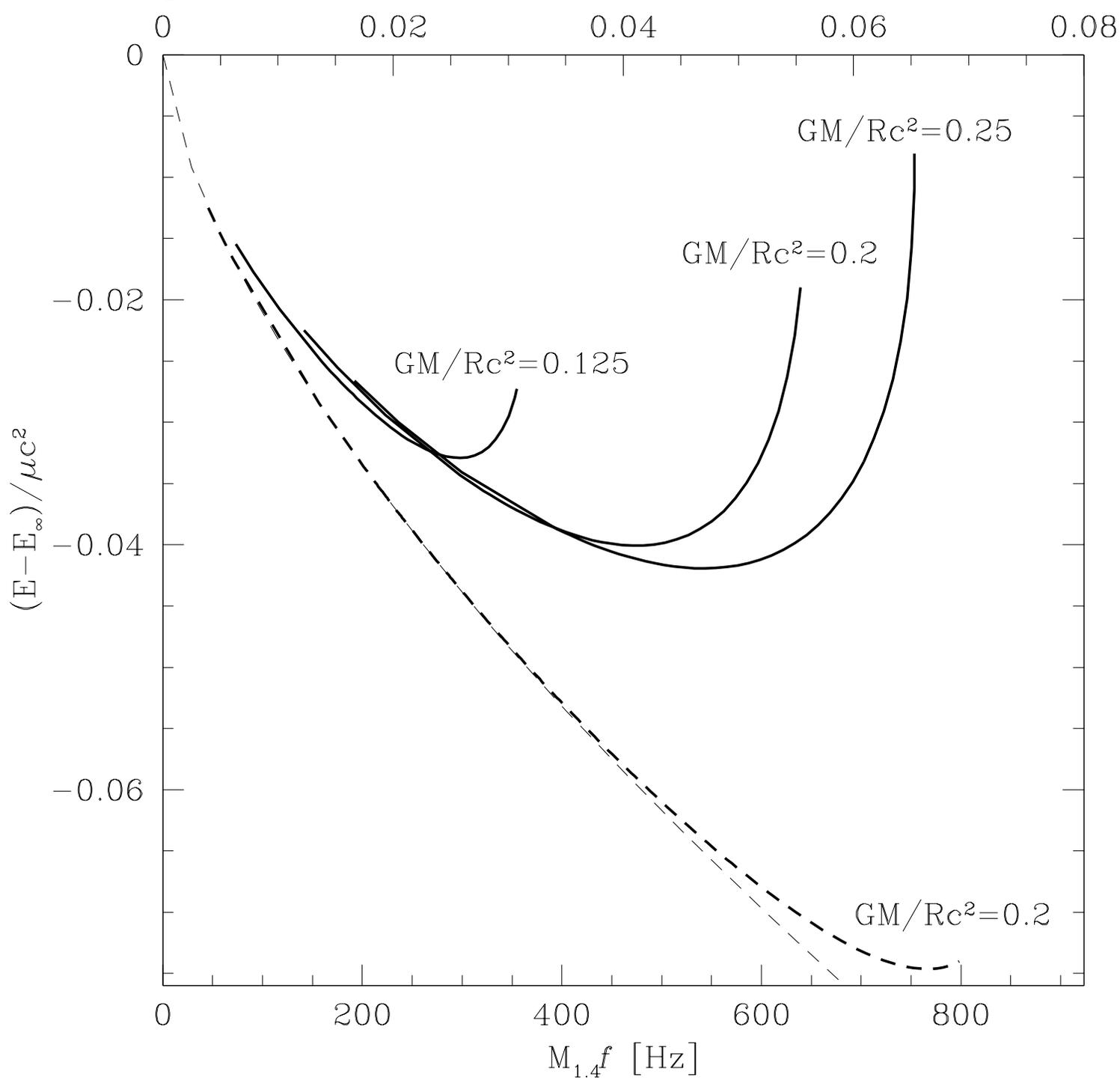
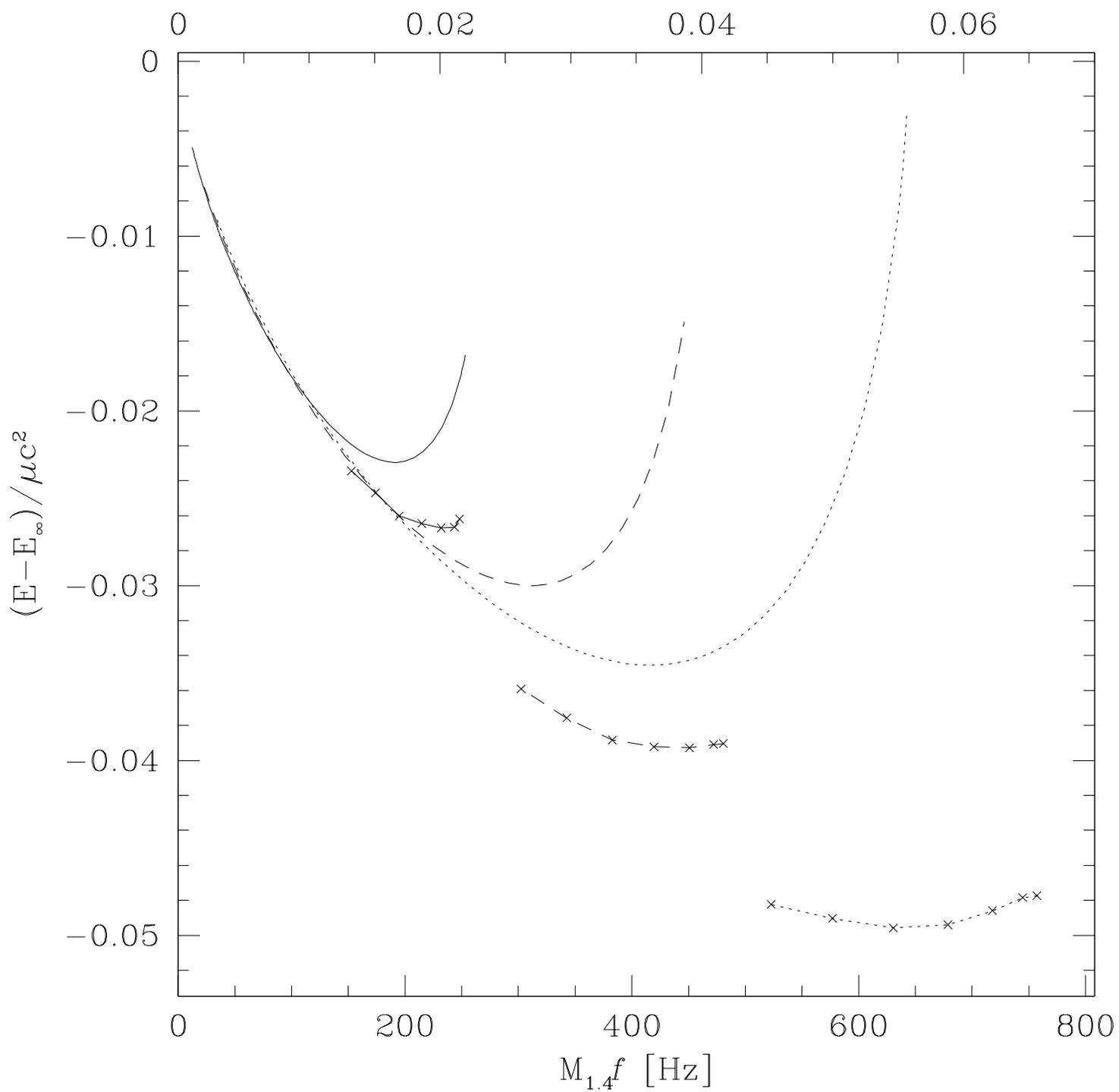
\end{document}